\documentclass{appolb}
\usepackage{epsfig}
\usepackage{graphicx}

\begin{document}
\title{New mechanisms of charm production.%
\thanks{Presented at \textit{Epihany 2012}, January 9-11, Cracow, Poland.}%
}
\author{Antoni Szczurek
\address{
The H. Niewodnicza\'{n}ski Institute of Nuclear Physics\\
Polish Academy of Sciences\\
ul. Radzikowskiego 152, 31-342 Krak\'ow, Poland; \\
Rzesz\'ow University,
ul. Rejtana 16A, 35-959 Rzesz\'ow, Poland.
}
}
\maketitle
\begin{abstract}
We discuss production of charm quarks, mesons as well as nonphotonic 
electrons in $pp$ scattering at RHIC.
The distributions in rapidity and transverse momentum
of charm and bottom quarks/antiquarks are calculated in 
the $k_t$-factorization approach.
The hadronization of heavy quarks is done by means of 
fenomenological fragmentation functions and
semileptonic decay functions are found by fitting
semileptonic decay data. 
Good description of the inclusive data at large transverse 
momenta of electrons is obtained and a missing strength 
at small transverse momenta of electrons is found.

In addition we discuss kinematical correlations between charged leptons 
from different mechanisms.
Reactions initiated by purely QED $\gamma^*\gamma^*$-fusion in elastic 
and inelastic $pp$ collisions as well as diffractive mechanism of 
exclusive $c \bar c$ production are included.
A good description of the dilepton invariant mass spectrum 
of the PHENIX collaboration is achieved. 
Distributions in the dilepton pair transverse 
momentum and in azimuthal angle 
between electron and positron are presented.

A new mechanism of exclusive production of $c \bar c$ is discussed.
Corresponding results are shown and the possibility of its
identification is discussed.

We discuss also production of two pairs of $c \bar c$ within a simple
formalism of double-parton scattering (DPS). Very large cross sections,
comparable to single-$c \bar c$ production, are predicted
for LHC energies.
Both total inclusive cross section as a function of energy and
differential distributions are shown.
We discuss a perspective how to identify the double scattering
contribution.
\end{abstract}


\section{Introduction}

In recent years the PHENIX and STAR collaborations
have measured transverse momentum distribution
of so-called nonphotonic electrons \cite{STAR_electrons,PHENIX_electrons}.
The dominant contribution
to the nonphotonic electrons/positrons comes from the
semileptonic decays of charm and/or beauty mesons.
Formally such processes can be divided into three subsequent stages.
First $c \bar c$ or $b \bar b$ quarks are produced.
The dominant mechanisms being gluon-gluon fusion at higher energies
or quark-antiquark annihilation close to the threshold.
Next the heavy quarks/antiquarks fragment into heavy
charmed mesons $D, D^*$ or $B, B^*$. The vector
$D^*$ and $B^*$ mesons decay strongly producing 
pseudoscalar $D$ and $B$ mesons.
Finally the heavy pseudoscalar mesons decay 
semileptonically producing electrons/positrons.


The hadronization of heavy quarks is usually done
with the help of phenomenological fragmentation functions
with parameters adjusted to the production of heavy mesons
in $e^+ e^-$ or $p \bar p$ collisions.

The last ingredient are semileptonic decays of heavy mesons.
Only recently the CLEO \cite{CLEO} and BABAR \cite{BABAR} 
collaborations has measured precise spectra of electrons/positrons 
coming from the decays of $D$ and $B$ mesons.
This is done by producing specific resonances: $\Psi(3770)$
which decays into $D$ and $\bar D$ mesons (CLEO) and
$\Upsilon(4S)$ which decays into $B$ and $\bar B$ mesons (BABAR).
In both cases the heavy mesons are almost at rest,
so in practice one measures the meson rest frame
distributions of electrons/positrons.

In this presentation the results for production of electrons have been 
obtained within the $k_t$-factorization approach. At relatively low
RHIC energies intermediate $x$-values of gluon distributions
become relevant.  The Kwiecinski unintegrated gluon
(parton) distributions seem the best suited in this context
\cite{Kwiecinski}.
We use both Peterson \cite{Peterson} and so-called perturbative 
\cite{BCFY95} fragmentation functions. The electron/positron decay 
functions fitted recently \cite{LMS09} to the recent CLEO and BABAR data
are used.

Recently the PHENIX collaboration has measured dilepton invariant mass
spectrum from $0$ to $8$ GeV in $pp$ collisions at $\sqrt{s}=200$ GeV 
\cite{PHENIX}.
Up to now, production of open charm and bottom was studied only in 
inclusive measurements of charmed mesons \cite{Tevatron_mesons} and 
electrons \cite{electrons} and only inclusive observables were
calculated in pQCD approach \cite{CNV05,LMS09}. 

Some time ago we have studied kinematical correlations of $c \bar c$
quarks \cite{LS06}, which is, however, difficult to study experimentally.
High luminosity and in a consequence better statistics at present
colliders opens a new possibility to study not only inclusive
distributions but also correlations between outgoing particles.
Kinematical correlations constitute an alternative method to pin down 
the cross section for charm and bottom production. 

In this presentation I shall show some selected results obtained in
\cite{LMS09,MSS11}. 

Recently we have studied a new mechanism of exclusive production
of $c \bar c$ pairs \cite{MPS2010}. In such a process a
single pair of $c \bar c$ is produced together with associated two protons.
We shall comment on a possibility to identify the peculiar mechanism.
The original presentation at the conference included
also inclusive diffractive processes.

With growing energy the heavy quark-antiquark production becomes
sensitive to lower-$x$ gluon distributions. At high energies a possibility
of two pairs of $c \bar c$ production opens up \cite{LMS11_DPS}.
We comment on which areas of the phase space are potentially
interesting in order to pin down the double parton scattering contribution.


\section{Inclusive production of $c \bar c$}

\subsection{Formalism}

We consider the reaction $h_1 + h_2 \to Q + \bar Q + X$,
where $Q$ and $\bar Q$ are heavy quark and heavy antiquark,
respectively.


\begin{figure}[!thb]
\begin{center}
\includegraphics[width=4.5cm]{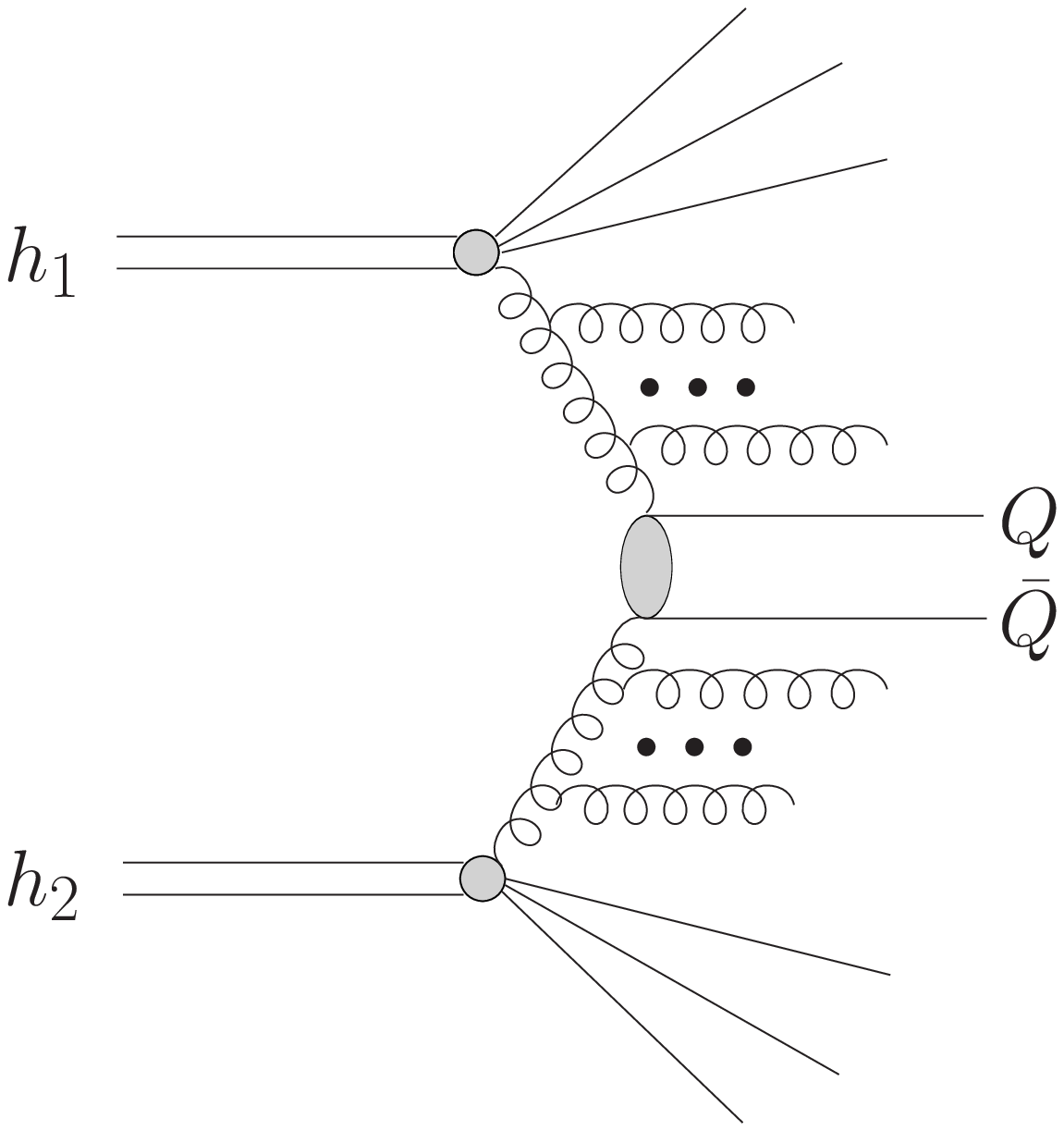}
\includegraphics[width=5.2cm]{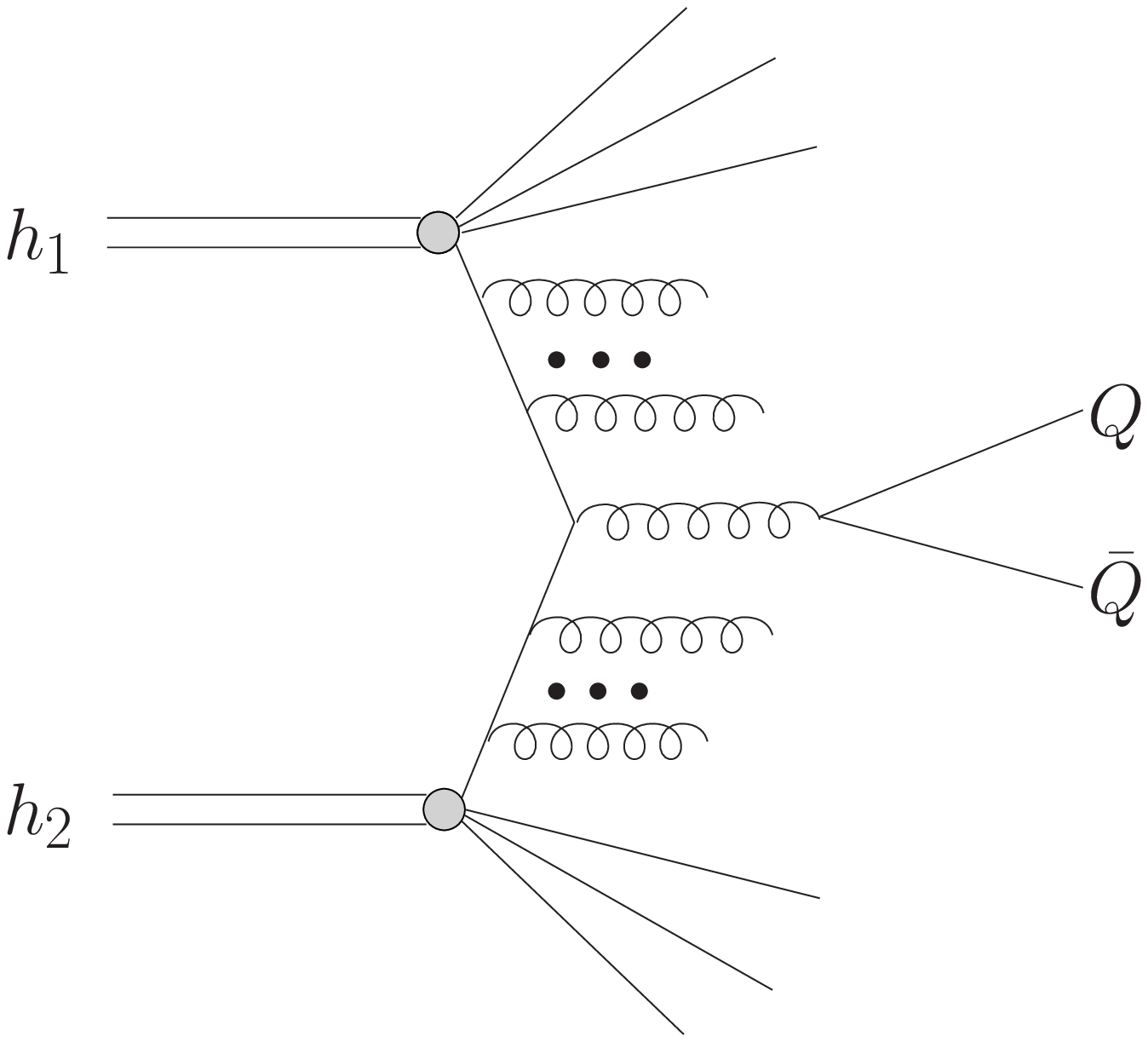}
\caption[*]{
\small Basic diagrams relevant for gluon-gluon fusion 
(left panel) and quark-antiquark annihilation (right panel)
in the $k_t$-factorization approach.
\label{fig:diagrams}
}
\end{center}
\end{figure}


%
%

In the $k_t$-factorization approach the differential cross
section reads:
\begin{eqnarray}
\frac{d \sigma}{d y_1 d y_2 d^2p_{1,t} d^2p_{2,t}} = \sum_{i,j} \;
\int \frac{d^2 \kappa_{1,t}}{\pi} \frac{d^2 \kappa_{2,t}}{\pi}
\frac{1}{16 \pi^2 (x_1 x_2 s)^2} \; \overline{ | {\cal M}_{ij} |^2}
\nonumber \\  
\delta^{2} \left( \vec{\kappa}_{1,t} + \vec{\kappa}_{2,t} 
                 - \vec{p}_{1,t} - \vec{p}_{2,t} \right) \;
{\cal F}_i(x_1,\kappa_{1,t}^2) \; {\cal F}_j(x_2,\kappa_{2,t}^2) \; ,
\label{LO_kt-factorization}    
\end{eqnarray}
where ${\cal F}_i(x_1,\kappa_{1,t}^2)$ and ${\cal F}_j(x_2,\kappa_{2,t}^2)$
are the so-called unintegrated gluon (parton) distributions.
Leading-order matrix elements for off-shell gluons
\cite{kt-factorization} 
were used.
The two-dimensional Dirac delta function assures momentum conservation.
The unintegrated parton distributions are evaluated at:\\
$x_1 = \frac{m_{1,t}}{\sqrt{s}}\exp( y_1) 
     + \frac{m_{2,t}}{\sqrt{s}}\exp( y_2)$,
$x_2 = \frac{m_{1,t}}{\sqrt{s}}\exp(-y_1) 
     + \frac{m_{2,t}}{\sqrt{s}}\exp(-y_2)$,\\
where $m_{i,t} = \sqrt{p_{i,t}^2 + m_Q^2}$.

Introducing new variables:
$\vec{Q}_t = \vec{\kappa}_{1,t} + \vec{\kappa}_{2,t} \; , \nonumber \\
\vec{q}_t = \vec{\kappa}_{1,t} - \vec{\kappa}_{2,t}$ 
one can write:
\begin{eqnarray}
\frac{d \sigma_{ij}}{d y_1 d y_2 d^2p_{1,t} d^2p_{2,t}} =
\int d^2 q_t \; \frac{1}{4 \pi^2}
\frac{1}{16 \pi^2 (x_1 x_2 s)^2} \; \overline{ | {\cal M}_{ij} |^2}
\nonumber \\  
{\cal F}_i(x_1,\kappa_{1,t}^2) \; {\cal F}_j(x_2,\kappa_{2,t}^2) \; .
\label{LO_kt-factorization2}    
\end{eqnarray}
This formula is very useful to study correlations between
the produced heavy quark $Q$ and heavy antiquark 
$\bar Q$ \cite{LS06}.

%
%

At the Tevatron and LHC energies the contribution of 
the $gg \to Q \bar Q$ subrocess is more than one order 
of magnitude larger than its counterpart for 
the $q \bar q \to Q \bar Q$ subprocess. At RHIC energy
the relative contribution of the $q \bar q$ 
annihilation is somewhat bigger.
Therefore in the following we shall take into account
not only gluon-gluon fusion process
but also the quark-antiquark annihilation mechanism.


The production of electrons/positrons is a multi-step
process.
The whole procedure of electron/positron production
can be written in the following schematic way:
\begin{equation}
\frac{d \sigma^e}{d y d^2 p} =
\frac{d \sigma^Q}{d y d^2 p} \otimes
D_{Q \to D} \otimes
f_{D \to e} \; ,
\label{whole_procedure}
\end{equation}
where the symbol $\otimes$ denotes a convolution of the different 
distributions.
The first term is responsible for production
of heavy quarks/antiquarks.
Next step is the process of formation of heavy mesons.
We follow a phenomenological approach and take e.g. Peterson 
\cite{Peterson} and Braaten et al. \cite{BCFY95} fragmentation functions
with parameters from the literature \cite{PDG_new}.
The electron decay function accounts for the proper branching fractions. 

The inclusive distributions of hadrons can be calculated as a
convolution of inclusive distributions of heavy quarks/antiquarks and 
Q $\to$ h fragmentation functions:
\begin{equation}
\frac{d \sigma (y_1, p_{1t}^{H}, y_{2}, p_{2t}^{H}, \phi)}{d y_1 d p_{1t}^{H} d y_{2} d p_{2t}^{H} d \phi}
 \approx
\int \frac{D_{Q \to H}(z_{1})}{z_{1}}\cdot \frac{D_{\bar Q \to \bar H}(z_{2})}{z_{2}}\cdot
\frac{d \sigma (y_1, p_{1t}^{Q}, y_{2}, p_{2t}^{Q}, \phi)}{d y_1 d
  p_{1t}^{Q} d y_{2} d p_{2t}^{Q} d \phi} d z_{1} d z_{2} \; ,
\end{equation}
where: 
{$p_{1t}^{Q} = \frac{p_{1t}^{H}}{z_{1}}$, $p_{2t}^{Q} =
  \frac{p_{2t}^{H}}{z_{2}}$, with
meson longitudinal fractions  $z_{1}, z_{2}\in (0,1)$.


We use decay functions fitted recently \cite{LMS09} to the CLEO and BABAR data.
In our approach the electrons (positrons) are
generated isotropically in the heavy meson rest frame.

%

\subsection{Results}

\subsubsection{Single electron spectra}

Before we start presenting our single particle distributions let us
focus on the decay functions discussed shortly above.
In Fig.\ref{fig:cleo_and_babar} we show our fit \cite{LMS09} to the
CLEO and BABAR data. The good quality fit of the data allows
to obtain reliable predictions for electron/positron single particle spectra.


\begin{figure}[!thb] %
\begin{center}
\includegraphics[width=5.0cm]{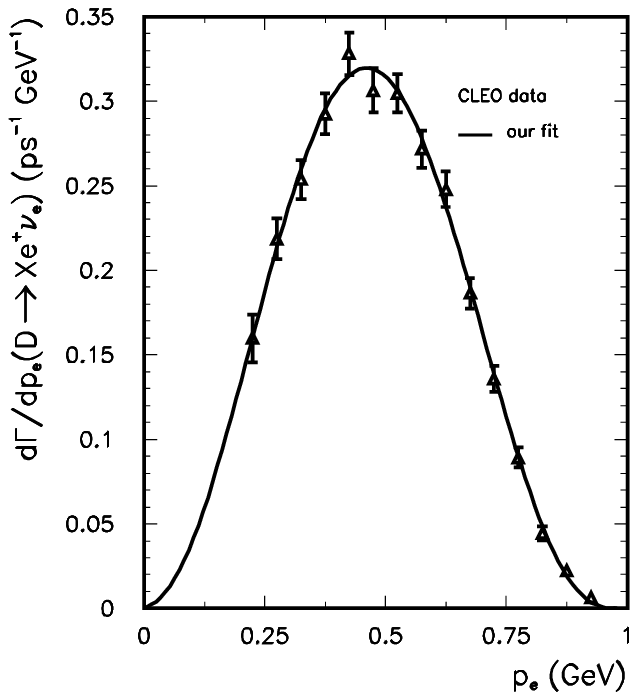}
\includegraphics[width=5.0cm]{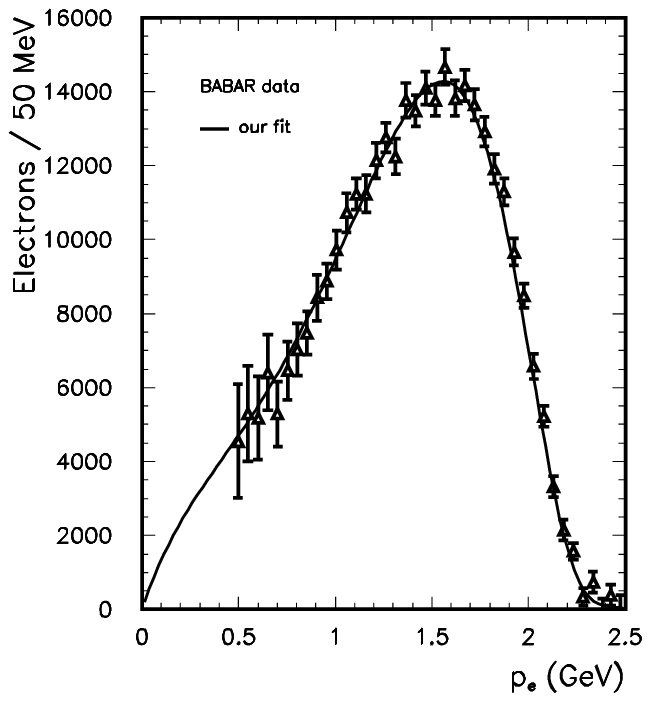}
\end{center}
\caption[*]{
\small Our fit to the CLEO \cite{CLEO} and BABAR \cite{BABAR}
data.
\label{fig:cleo_and_babar}
}
\end{figure}




Now we shall focus on transverse momentum distribution
of electrons/positrons measured recently by
the STAR and PHENIX collaborations at RHIC 
\cite{STAR_electrons,PHENIX_electrons}.
In Fig.\ref{fig:dsig_dpt_kwiecinski1}, as an example,
we show results obtained with the Kwieci\'nski UPDFs \cite{Kwiecinski}. 
In Ref.\cite{LMS09} we have discussed in addition other UGDFs.
Uncertainties due to different combinations of factorization and 
renormalization scales as well as due to different choices
fragmentation functions
are shown in Fig.\ref{fig:dsig_dpt_kwiecinski_scale_uncertainty_band}.
In these calculations we have included both
gluon-gluon fusion as well as quark-antiquark annihilation.
In the last case we use matrix elements with on-shell
formula but for off-shell kinematics (the discussion
of this point can be found in our earlier paper
\cite{LS06}).
In Ref.\cite{LMS09} we have discussed also uncertainties due to the choice
of quark masses.

\begin{figure}[!thb] %
\begin{center}
\includegraphics[width=5cm]{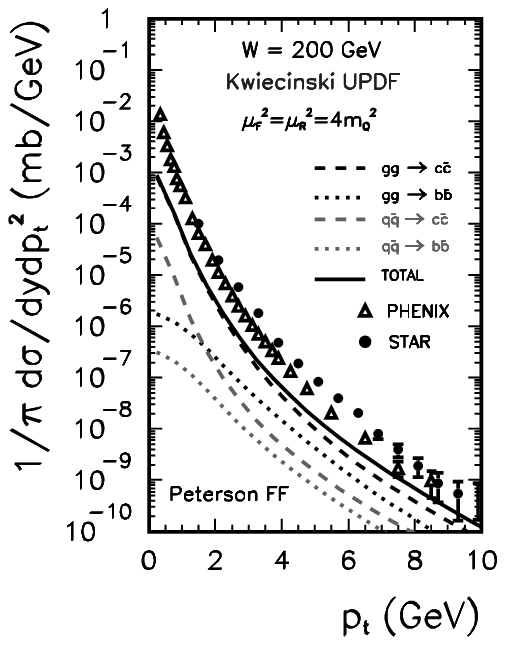}
\includegraphics[width=5cm]{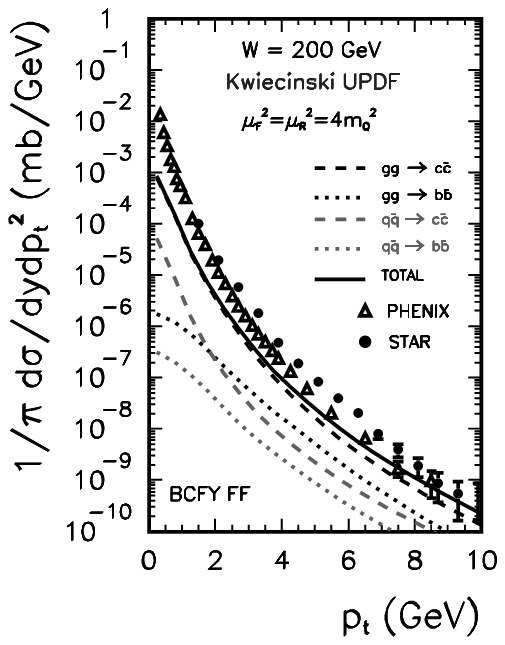}
\end{center}
\caption[*]{
\small Transverse momentum distribution of electrons/positrons
obtained with the Kwieci\'nski UPDFs. 
We show separately contributions of the gluon-gluon
fusion (black) and quark-antiquark annihilation (grey).
On the left side results with the Peterson
fragmentation functions and on the right side with
BCFY fragmentation functions. 
\label{fig:dsig_dpt_kwiecinski1}
}
\end{figure}








Study of nonphotonic $e^{\pm}$ and hadron correlations
allows to "extract" a fractional contribution
of the bottom mesons $B / (D + B)$ as a function
of electron/positron transverse momentum 
\cite{STAR_B_to_DB}. Recently the STAR collaboration
has extended the measurement of the relative
$B$ contribution to electron/positron transverse momenta
$\sim$ 10 GeV \cite{Mischke08a}.
In Fig.\ref{fig:B_fraction_Kwiecinski_scales} 
(Kwieci\'nski UPDFs) shown are results 
for different scales and different fragmentation functions.
There is a strong dependence on the factorization
and renormalization scale.
A slightly better agreement is obtained with the Peterson
fragmentation functions. 



\begin{figure}[!thb] %
\begin{center}
\includegraphics[width=5.0cm]{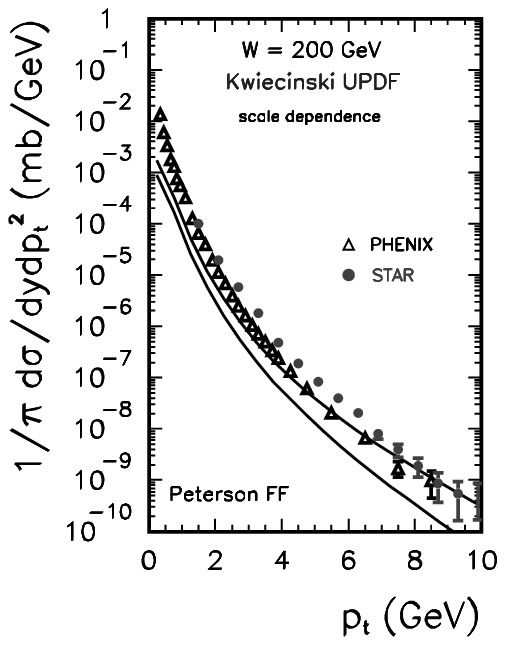}
\includegraphics[width=5.0cm]{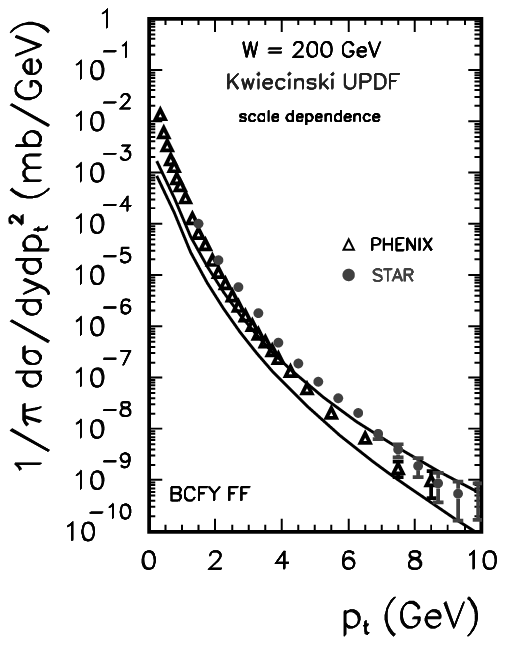}
\end{center}
\caption[*]{
\small Factorization and renormalization uncertainty band of 
our $k_t$-factorization calculation 
with unintegrated Kwieci\'nski gluon, quark and antiquark
distributions for the Peterson fragmentation function 
(left panel) and BCFY fragmentation function (right panel).
The open triangles represent the PHENIX collaboration data
and the solid circles the STAR collaboration data.
\label{fig:dsig_dpt_kwiecinski_scale_uncertainty_band}
}
\end{figure}






\begin{figure}[!thb] %
\begin{center}
\includegraphics[width=5.0cm]{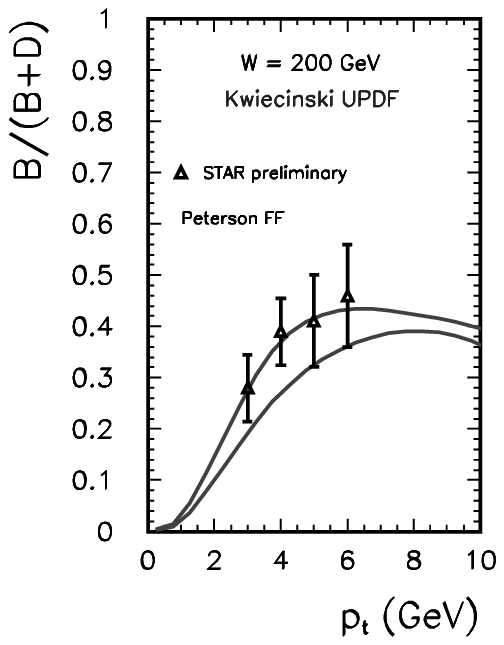}
\includegraphics[width=5.0cm]{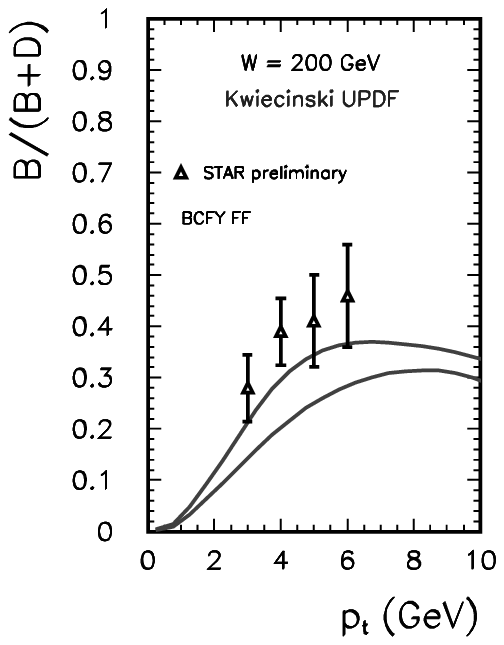}
\end{center}
\caption[*]{
\small The fraction of the B decays for the Kwieci\'nski
UPDFs. The uncertainty band due to the choice of the scales
is shown for the Peterson (left) and Braaten et al. (right)
fragmentation functions. Both gluon-gluon fusion as well
as quark-antiquark annihilation are included in this 
calculation.
\label{fig:B_fraction_Kwiecinski_scales}
}
\end{figure}








\subsubsection{Electron-positron correlations}

When calculating correlation observables we have included also 
processes shown in Fig.\ref{fig:qed} and Fig.\ref{fig:CED_mechanism}.
The photon-photon induced processes were first included in
Ref. \cite{MSS11}.
The central exclusive diffractive process shown in 
Fig.\ref{fig:CED_mechanism} was proposed in Ref.\cite{MPS2010}.
 
\begin{figure}[!h]
 \includegraphics[width=3.0cm]{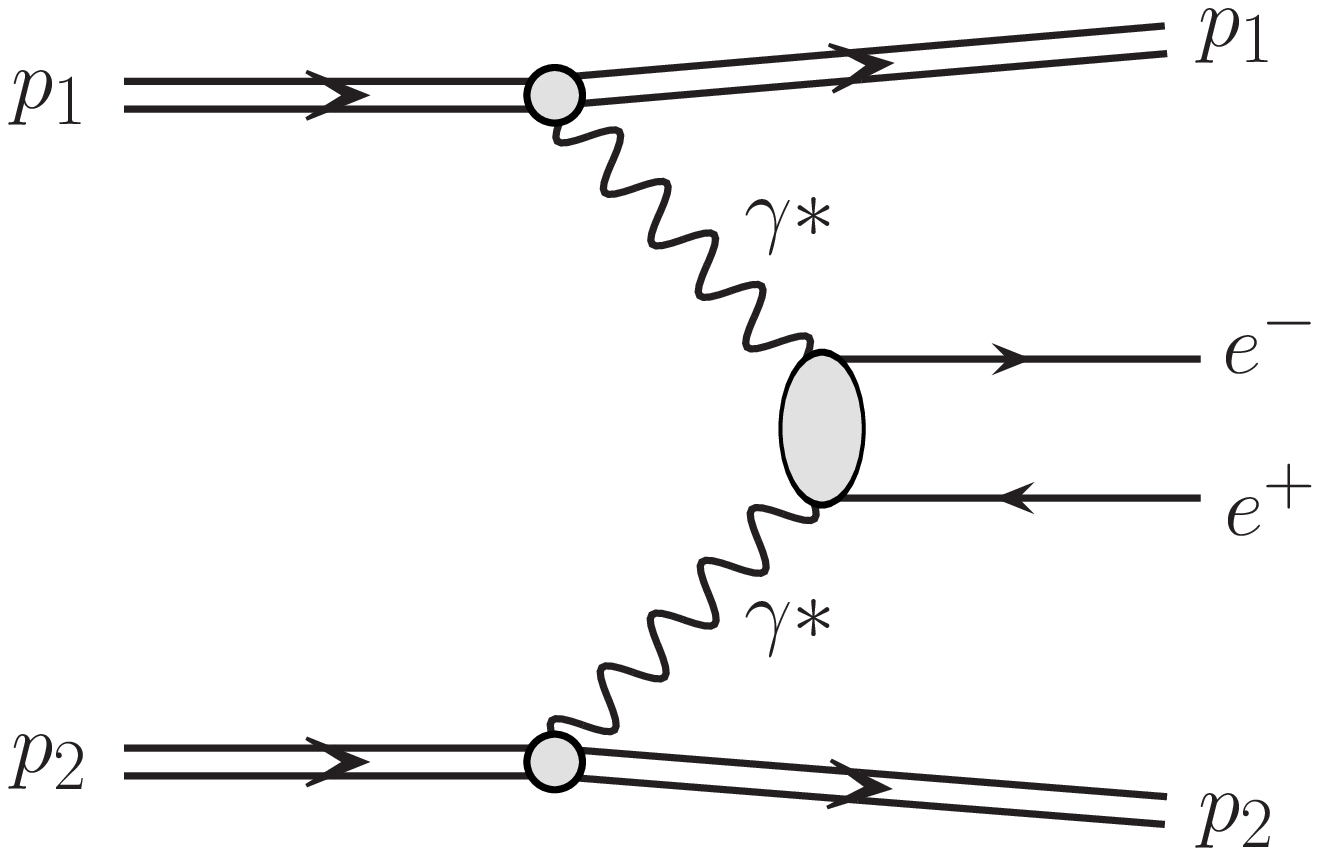}
 \includegraphics[width=3.0cm]{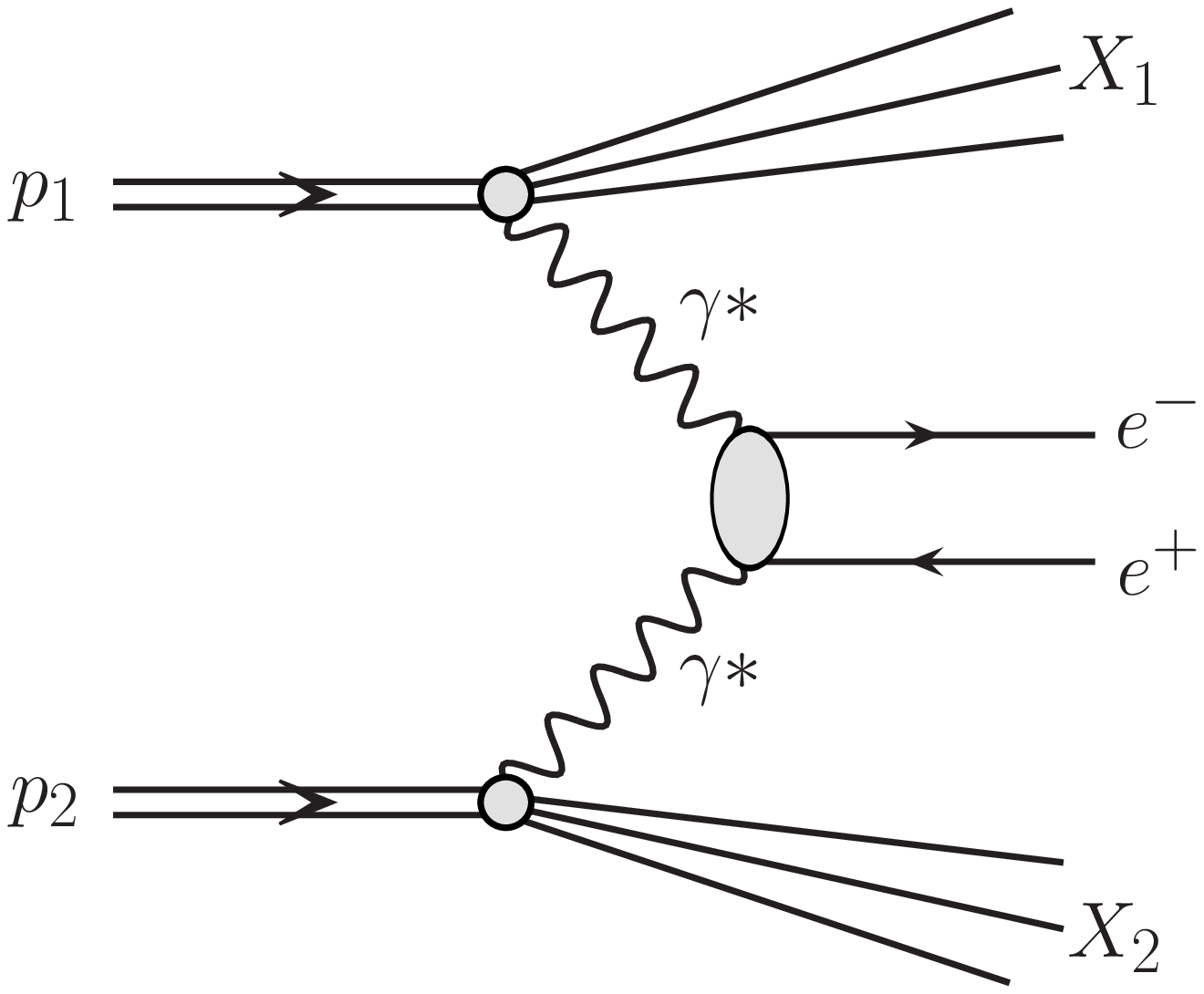}
 \includegraphics[width=3.0cm]{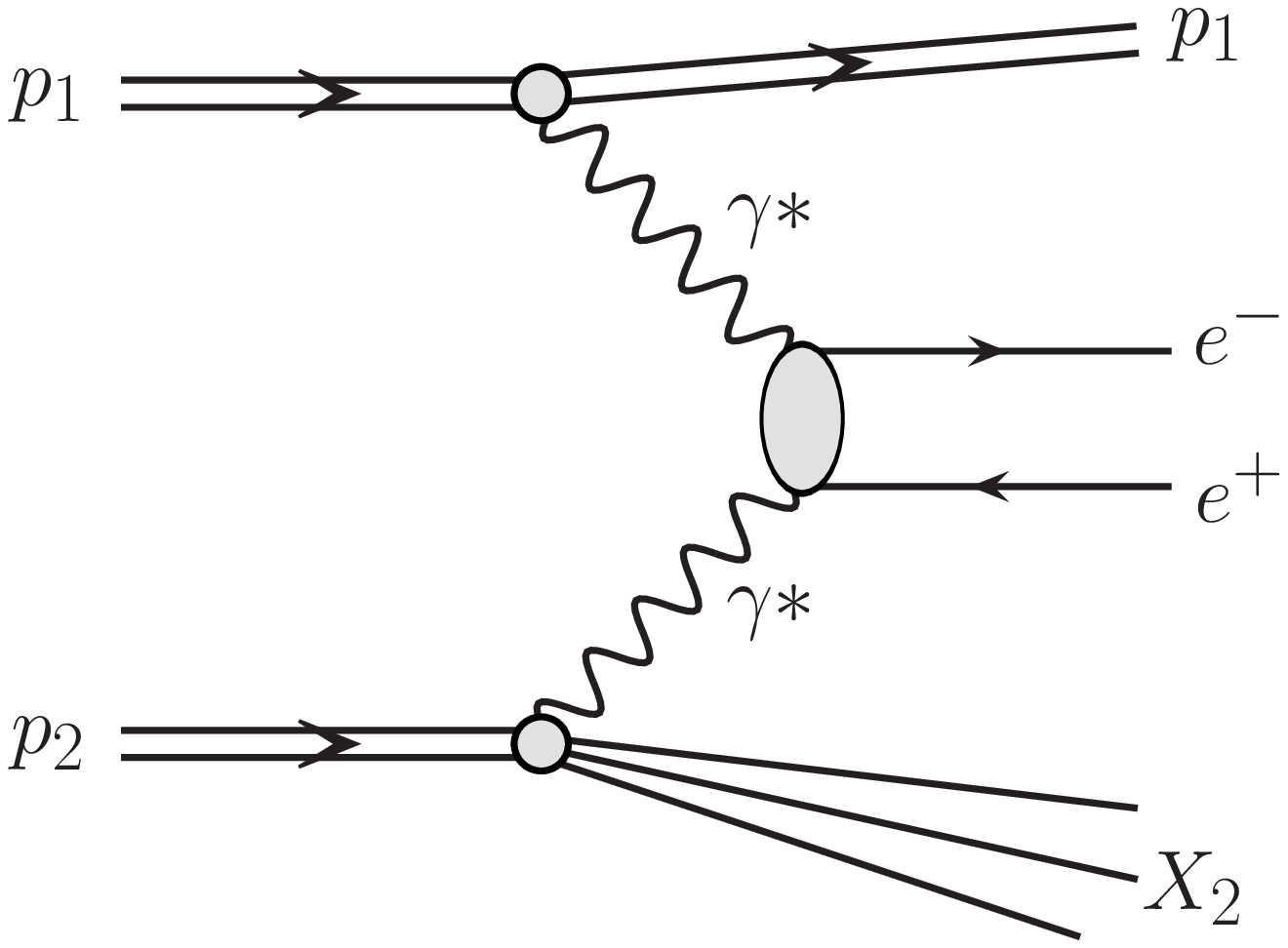}
 \includegraphics[width=3.0cm]{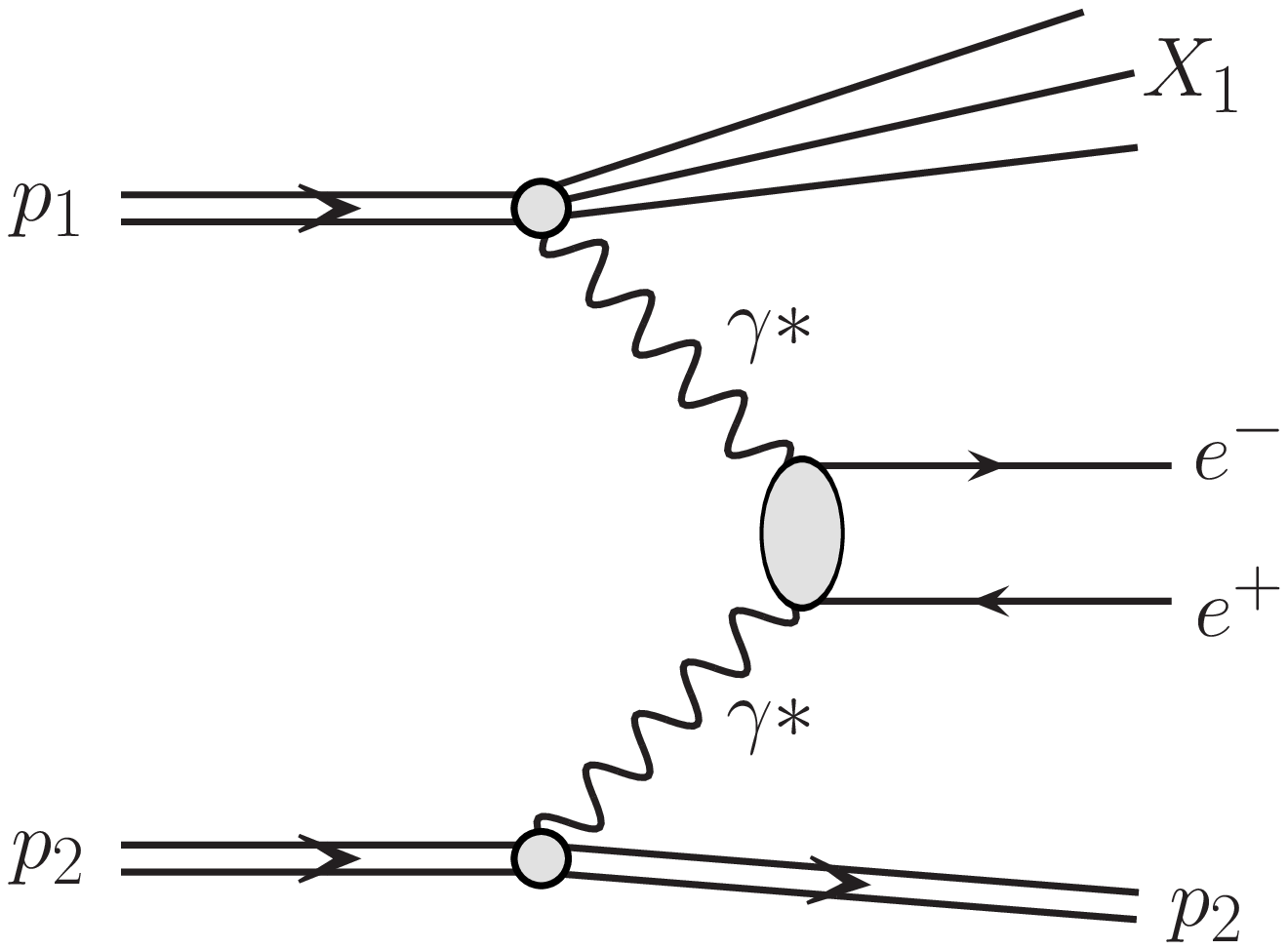}
   \caption{
\small Diagrammatic representation of processes initiated by
photon-photon subprocesses: double-elastic, 
double-inelastic, inelastic-elastic
and elastic-inelastic.
}
 \label{fig:qed}
\end{figure}

\begin{figure}[h!]
\begin{center}
\includegraphics[width=5.0cm]{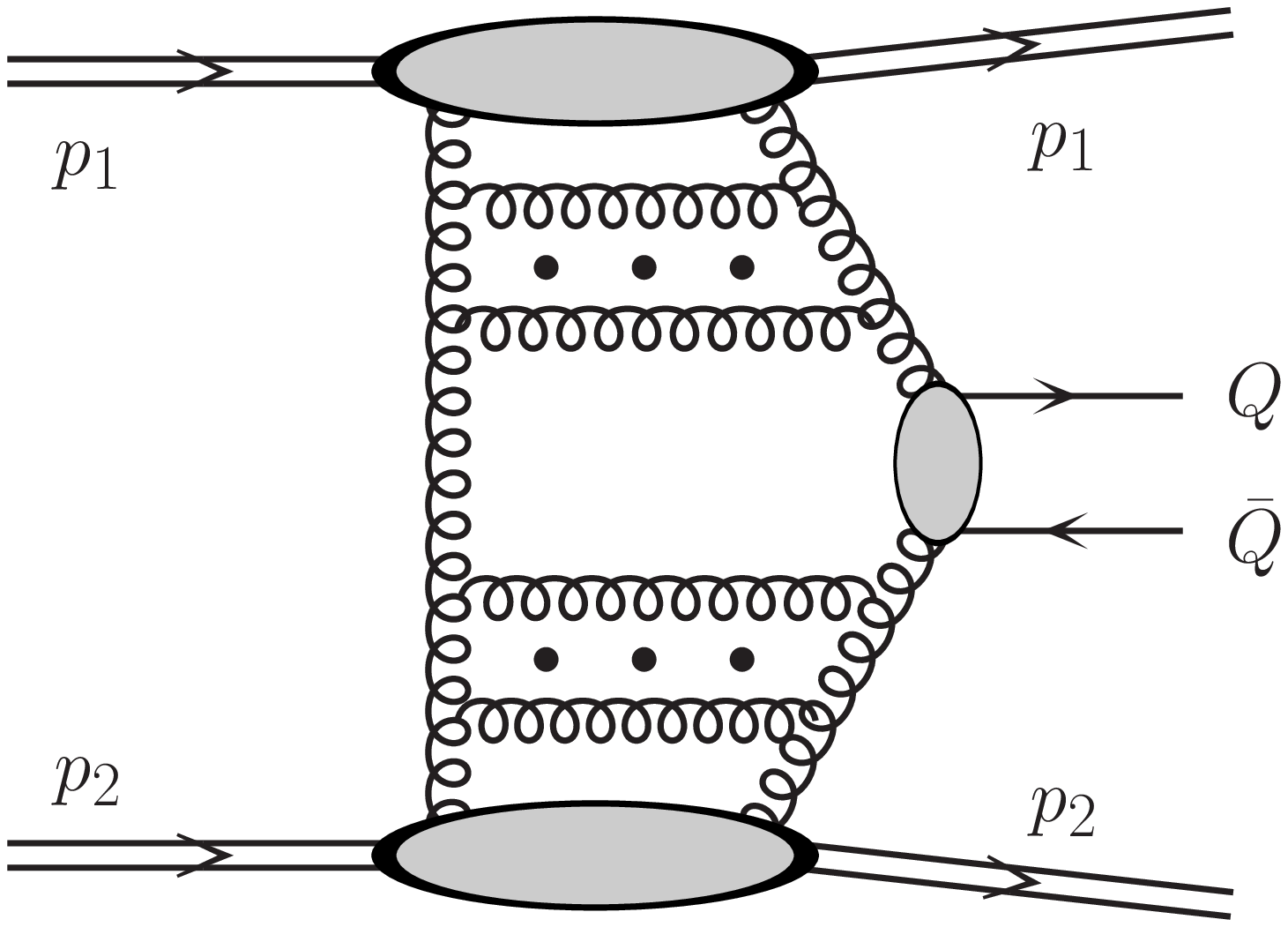}
\end{center}
\caption{\small The mechanism of exclusive double-diffractive
production of open charm.}
\label{fig:CED_mechanism}
\end{figure}

In Fig.\ref{fig:dsig_dM} we show $e^+  e^-$ invariant mass distributions
calculated with the Kwiecinski (left) and KMR (right) UGDFs. One can
clearly see that both the Kwiecinski and KMR \cite{KMRupdf} UGDFs give
fairly good description of the data for $M_{e^{+}e^{-}} >$ 3 GeV. At small 
invariant masses the Kwiecinski UGDF underestimates the PHENIX data and 
the KMR UGDF starts to overestimate the data points below 
$M_{e^{+}e^{-}}$ = 2 GeV.

\begin{figure}[!h]
\begin{center}
\includegraphics[width=5.0cm]{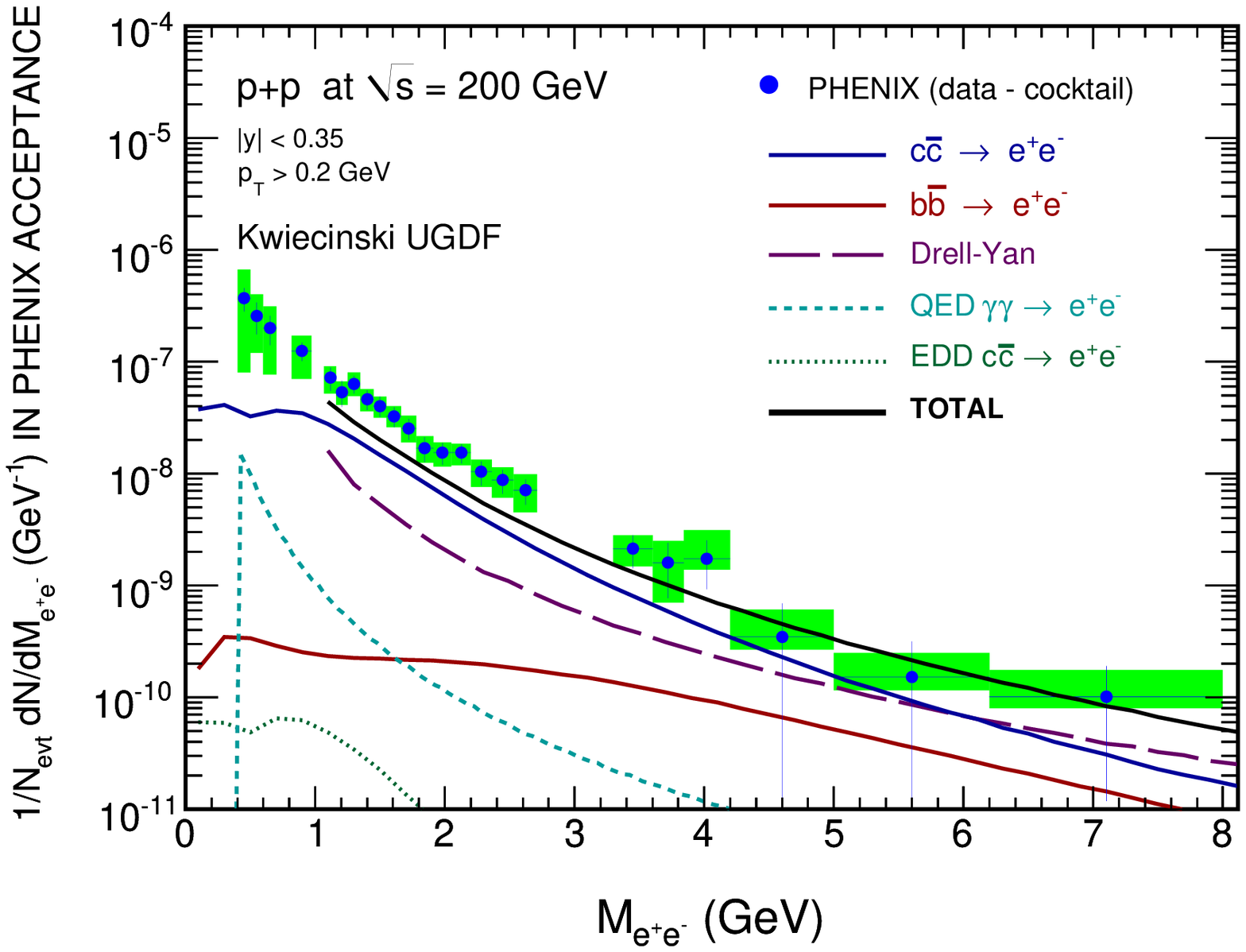}
\includegraphics[width=5.0cm]{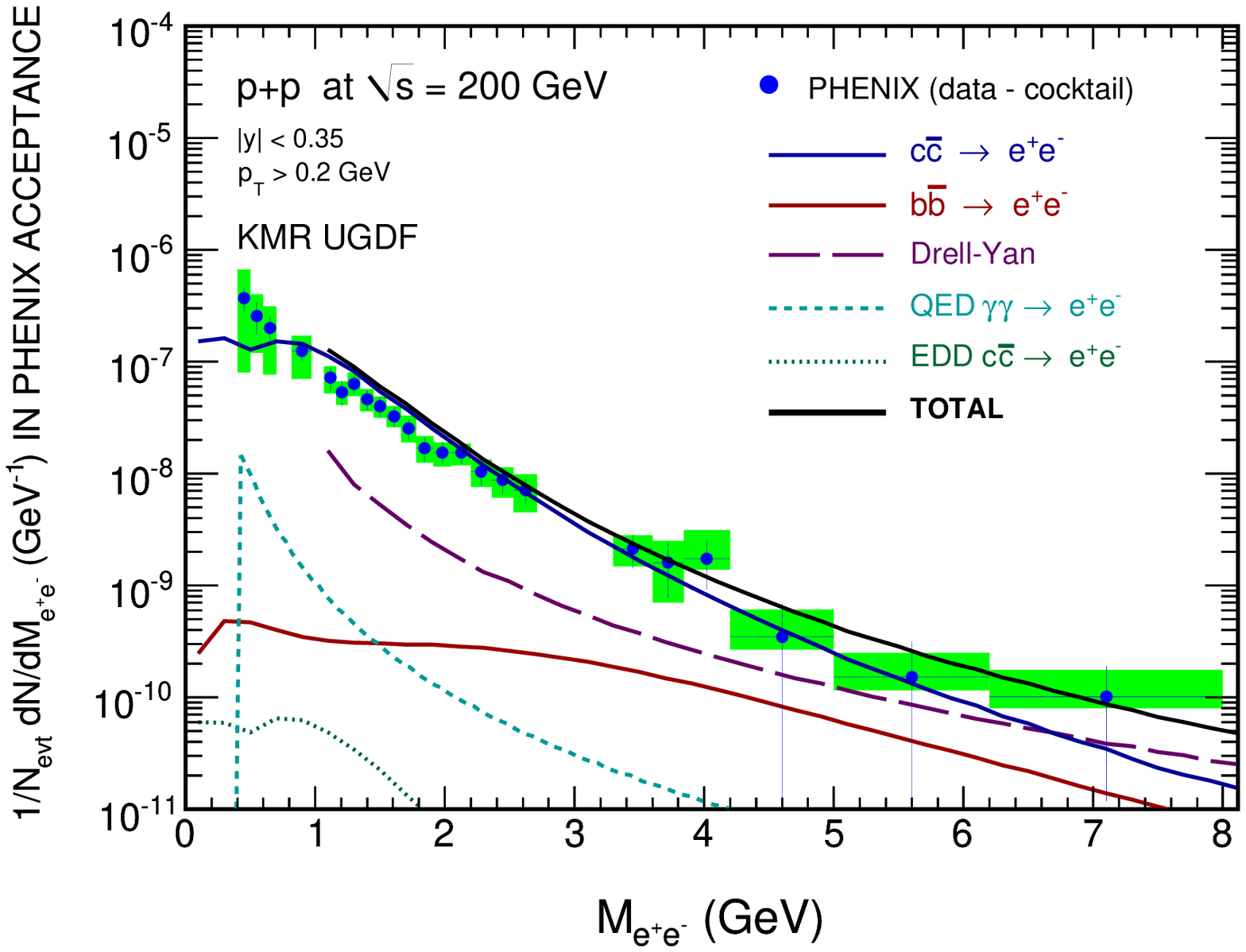}
\end{center}
   \caption{
\small Dielectron invariant mass distribution for $pp$
collisions at $\sqrt{s}$ = 200 GeV for the Kwieci\'nski (left) and KMR
(right) UGDFs. Different contributions are
shown separately: semileptonic decay of charm by the blue solid line, 
semileptonic decay of bottom by the red solid line, Drell-Yan mechanism by
the long dashed line, gamma-gamma processes by the blue dashed line and 
the central diffractive contribution by the green dotted line.
In this calculation we have included azimuthal angle acceptance of the
PHENIX detector \cite{PHENIX}.
}
 \label{fig:dsig_dM}
\end{figure} 




In Fig.\ref{fig:uncertainties} we show uncertainties related to
the contribution of semileptonic decays. The left panel 
presents uncertainties due to the factorization scale variation as described
in the figure caption.
The right panel shows uncertainties due to the 
modification of the heavy quark masses.

\begin{figure}[!h]
\begin{center}
 \includegraphics[width=5.0cm]{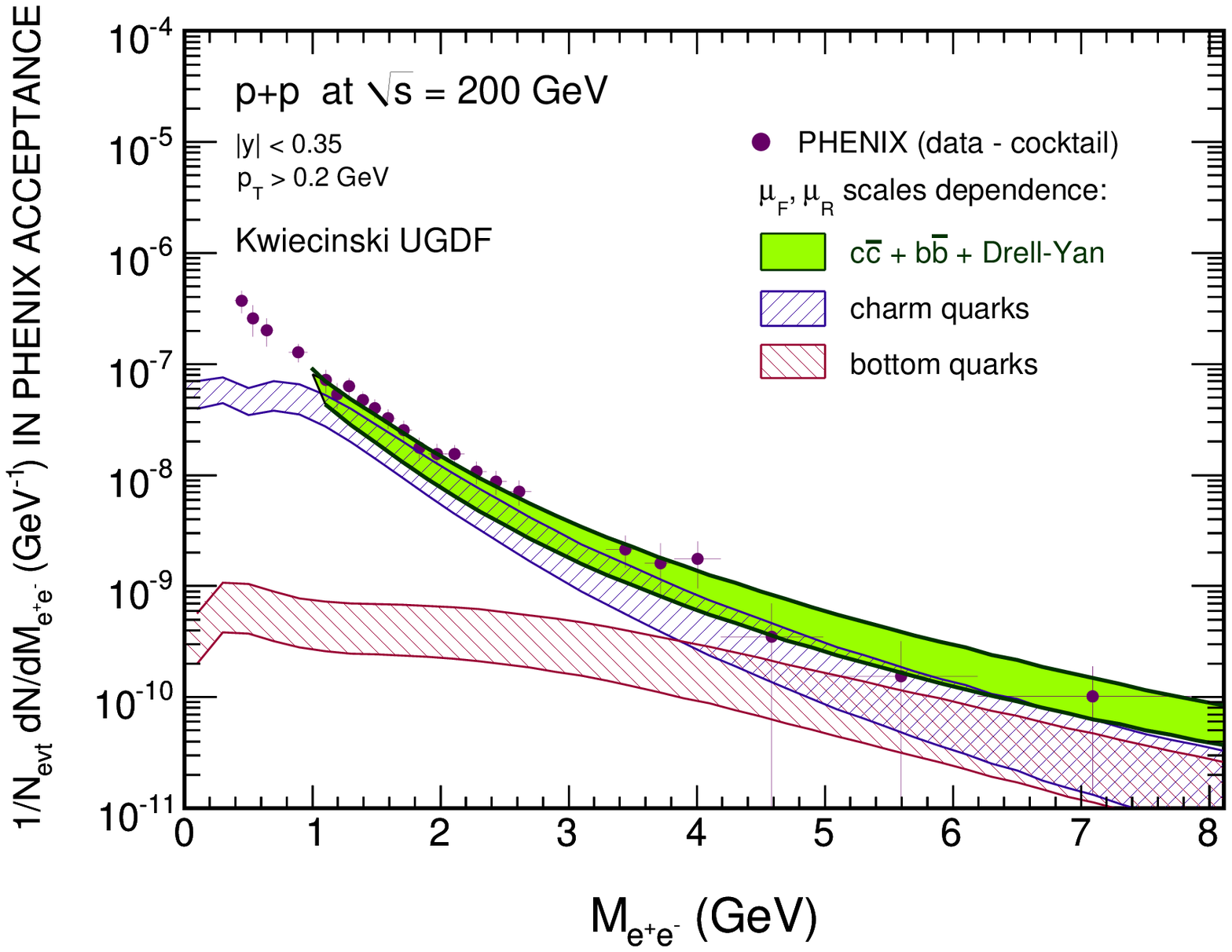}
 \includegraphics[width=5.0cm]{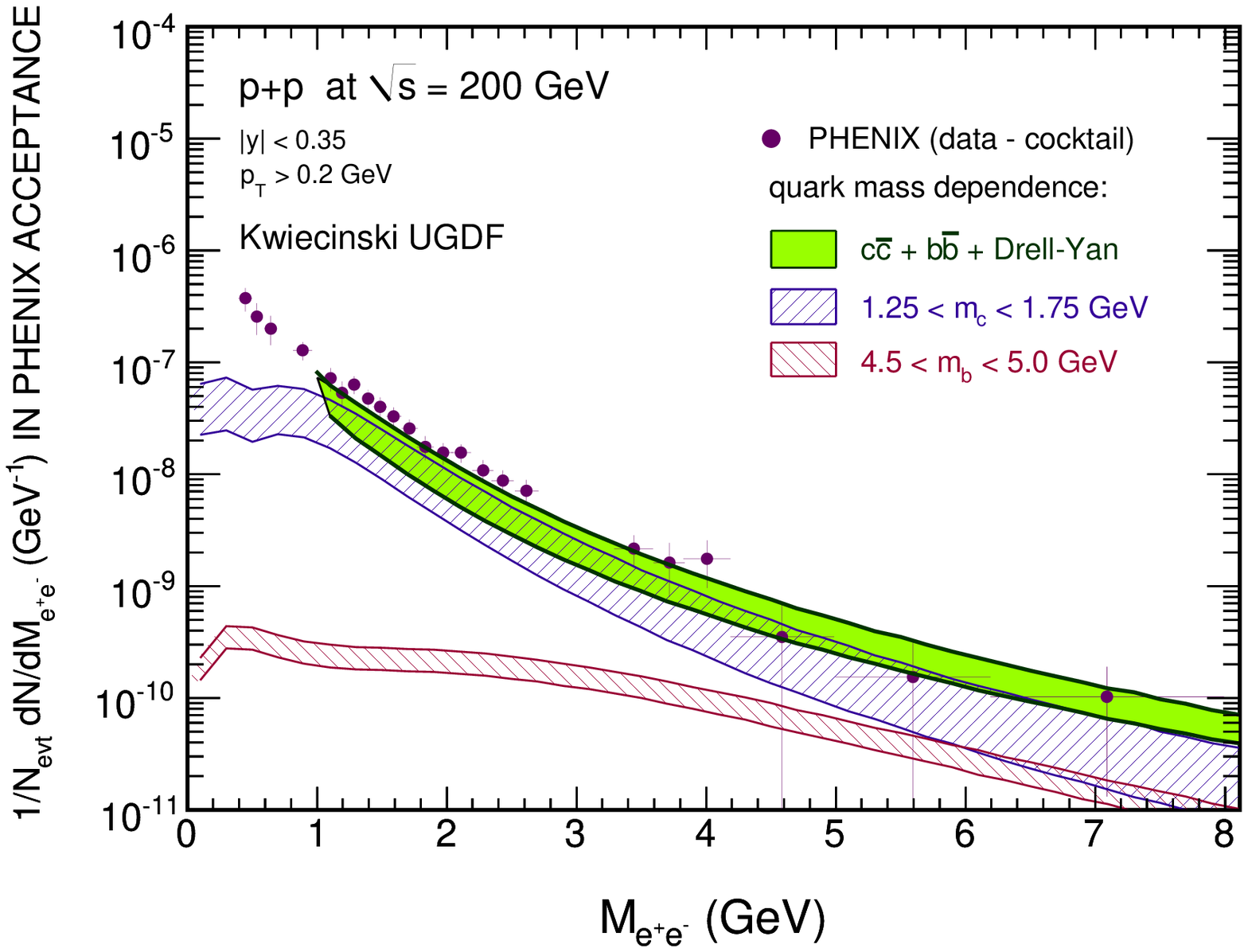}
\end{center}
   \caption{
\small The uncertainties of theoretical calculations.
The left panel shows the factorization scale uncertainties, 
the lower curve corresponds to $\mu_F^2, \mu_R^2 = m_{1,t}^2 + m_{2,t}^2$ and 
the upper curve to $\mu_R^2 = k_t^2$, $\mu_F^2 = 4 m_Q^2$, where $k_t$ 
is gluon transverse momentum.
The right panel shows the quark mass uncertainties as indicated
in the figure.
}
 \label{fig:uncertainties}
\end{figure}



If the detector can measure both transverse momenta of electron/positron
and their directions,
one can construct a distribution in transverse momentum of the dielectron
pair: $\vec{p}_{t,sum} = \vec{p}_{1t} + \vec{p}_{2t}$.
Our predictions including the semileptonic decays and Drell-Yan processes
are shown in the left panel of Fig.\ref{fig:new_distributions}. 
Both processes give rather similar distributions.
The distributions of this type were not measured so far experimentally.
The distribution in $p_{t,sum}$ is not only a consequence
of gluon transverse momenta 
but invlolves also fragmentation process and semileptonic decays. 
With good azimuthal resolution of detectors one could also construct
distribution in azimuthal angle between electron and positron. 
Corresponding predictions are shown in the right panel of 
Fig.\ref{fig:new_distributions}. One can see 
an interesting dependence on the invariant mass of the dielectron pair
-- the smaller the invariant mass the large the decorrelation in azimuthal
angle.

\begin{figure}[!h]
\begin{center}
\includegraphics[width=5.0cm]{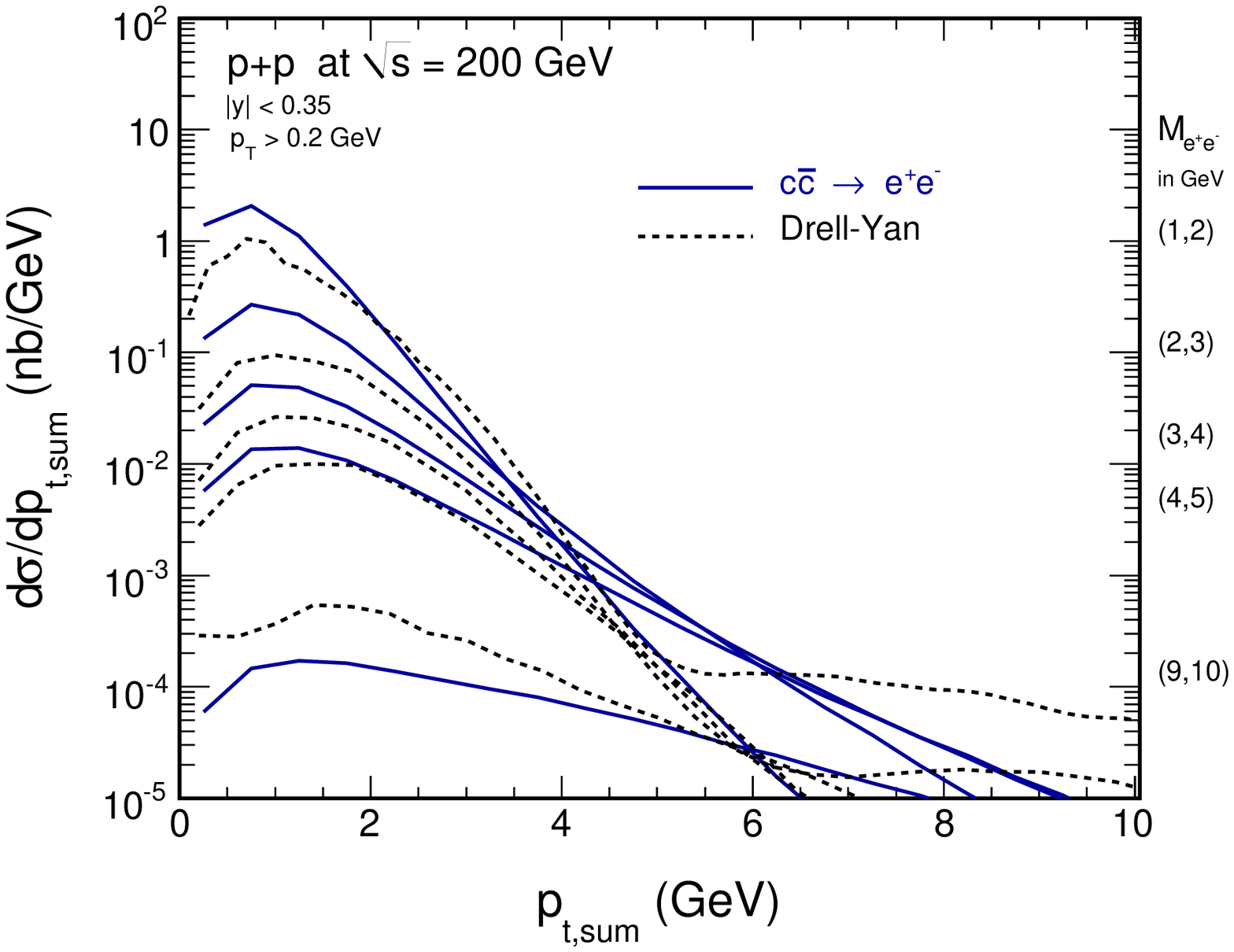}
\includegraphics[width=5.0cm]{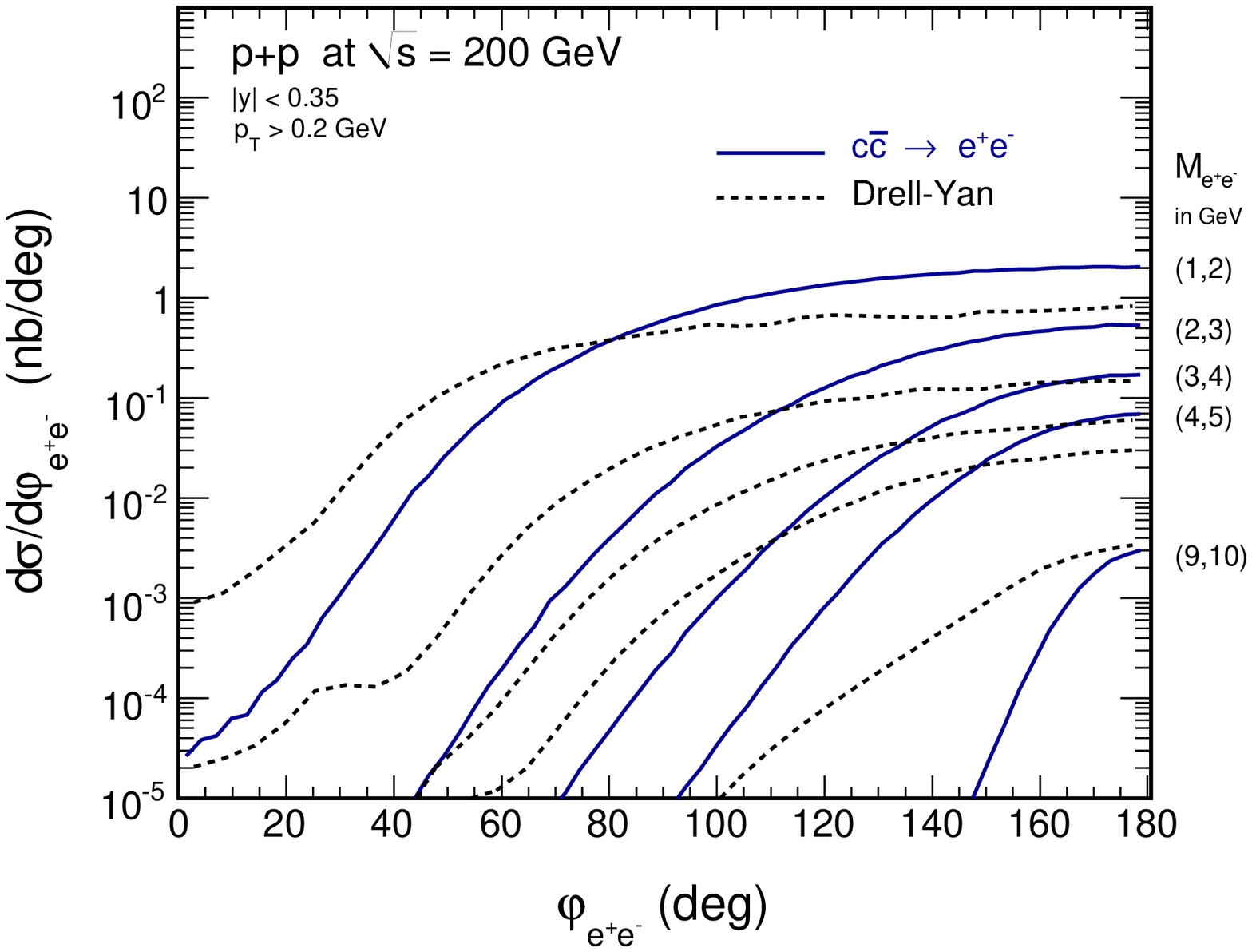}
\end{center}
   \caption{
\small
Distribution in transverse momentum of the dielectron pair (left)
and in azimuthal angle between electron and positron (right) 
for semileptonic decays (solid line) 
and Drell-Yan processes (dashed line). Here Kwiecinski UGDF and Peterson
fragmentation function were used.
}
 \label{fig:new_distributions}
\end{figure}


\section{Exclusive diffractive production of $c \bar c$}

Central exclusive mechanisms of $c \bar c$ production 
at high energies shown in Fig.\ref{fig:CED_mechanism} constitutes a special
category of diffractive processes.
In this case the central system $X$ is produced in the color singlet
state. This leads to rapidity gaps between forward/backward produced 
protons and the central system.
The QCD mechanism of central exclusive heavy quark-atiquark dijets
(in particularly $b \bar b$) is a source of the irreducible background
to the exclusive Higgs boson production \cite{our-bb,our-higgs}.
Central exclusive production of $c \bar c$ and $b \bar b$ pairs was 
studied in detail in our previous papers \cite{MPS2010, our-bb,
our-higgs}. In these calculations the $pp\rightarrow p (q \bar q) p$
reaction was considered 
as a 4-body process with exact kinematics. The applied
perturbative model of theoretical predictions is based on the 
Khoze-Martin-Ryskin (KMR) approach used previously for the exclusive
Higgs boson production \cite{KMR_Higgs}. 
Total cross sections and differential distributions for heavy quarks are
calculated by using $k_{t}$-factorization approach with the help of 
off-diagonal unintegrated gluon distribution functions.


This QCD model works very well in the case of exclusive dijets and 
charmonia production \cite{DKRS2011,MPS_dijets,PST_chic0,PST_chic12}. 
Here we discuss the production of 
$c \bar c$ pairs. In practice, however, one measures rather charmed
mesons. The measurement and its interpretation is therefore more complicated
and will be not discussed here.
Such experimental analyses are being performed now at the Tevatron and 
could be also available in Run II at RHIC.
In this context it is very interesting to compare the mechanism of central 
exclusive production of charm quarks with standard inclusive single and 
central diffractive processes.

As in the KMR approach 
\cite{KMR_Higgs, KMR-bb}
the amplitude of the exclusive central diffractive $q\bar{q}$ pair 
production $pp\to p(q\bar{q})p$ can be written as
\begin{eqnarray}
{\cal M}_{\lambda_q\lambda_{\bar{q}}}= &&
\frac{s}{2}\cdot\frac{\pi^2\delta_{c_1c_2}}{N_c^2-1}\, \Im\int d^2
q_{0,t} \; V_{\lambda_q\lambda_{\bar{q}}}^{c_1c_2}(q_1, q_2, k_1, k_2) 
\nonumber \\ 
&&\times\frac{f^{\mathrm{off}}_{g,1}(x_1,x_1',q_{0,t}^2,
q_{1,t}^2,t_1)f^{\mathrm{off}}_{g,2}(x_2,x_2',q_{0,t}^2,q_{2,t}^2,t_2)}
{q_{0,t}^2\,q_{1,t}^2\, q_{2,t}^2}, \label{amplitude}
\end{eqnarray}
where $\lambda_q,\,\lambda_{\bar{q}}$ are helicities of heavy $q$
and $\bar{q}$, respectively, $t_{1,2}$ are the momentum transfers
along each proton line, $q_{1,t}, q_{2,t}, x_{1,2}$ and
$q_{0,t},\,x_1'\sim x_2'\ll x_{1,2}$ are the transverse momenta and
the longitudinal momentum fractions for active and screening gluons,
respectively. Above $f^{\mathrm{off}}_{g,1/2}$ are the off-diagonal
UGDFs related to both nucleons. The vertex factor
$V_{\lambda_q\lambda_{\bar{q}}}^{c_1c_2}(q_1, q_2, k_1, k_2)$
is the production amplitude of a pair of massive quark $q$ and antiquark
$\bar q$ with helicities $\lambda_q$, $\lambda_{\bar{q}}$ and momenta 
$k_1$, $k_2$, respectively.
The longitudinal momentum fractions of active gluons
are calculated based on kinematical variables of outgoing quark
and antiquark:
$x_1 = \frac{m_{q,t}}{\sqrt{s}} \exp(+y_q)
     +  \frac{m_{\bar q,t}}{\sqrt{s}} \exp(+y_{\bar q})$ and
$x_2 = \frac{m_{q,t}}{\sqrt{s}} \exp(-y_q)
     +  \frac{m_{\bar q,t}}{\sqrt{s}} \exp(-y_{\bar q})$,
where $m_{q,t}$ and $m_{\bar q,t}$ are transverse masses of the quark and
antiquark, respectively, and $y_q$ and $y_{\bar q}$ are corresponding
rapidities.

The off-diagonal UGDFs can be approximated as \cite{KMR-ugdf}
 \begin{equation}
 f^{\mathrm{off}}_g(x',x_{1,2},q_{1,2t}^2,q_{0,t}^2,\mu_F^2)\simeq
 R_g\,f_g(x_{1,2},q_{1,2t}^2,\mu_F^2) \; .
 \label{rg}
 \end{equation}
%
The factor $R_g$ here cannot be calculated from first principles in 
the most general case of off-diagonal UGDFs. It can be estimated only in
the case of off-diagonal collinear PDFs
when $x' \ll x$ and $x g = x^{-\lambda}(1-x)^n$ and then $R_g = \frac{2^{2\lambda+3}}{\sqrt{\pi}}
\frac{\Gamma(\lambda+5/2)}{\Gamma(\lambda+4)}$. In the considered kinematics
the diagonal unintegrated densities can be written in terms of the
conventional (integrated) densities $xg(x,q_t^2)$ as~\cite{KMR-ugdf}
\begin{equation}\label{ugdfkmr}
f_g(x,q_t^2,\mu^2)=\frac{\partial}{\partial\ln q_t^2}
\big[xg(x,q_t^2)\sqrt{T_g(q_t^2,\mu^2)}\big] \; ,
\end{equation}
where $T_g$ is the conventional Sudakov survival factor which
suppresses real emissions from the active gluon during the
evolution.

The hard subprocess $g^*g^*\to q\bar q$
amplitude $V_{\lambda_q\lambda_{\bar{q}}}^{c_1c_2}(q_1, q_2, k_1, k_2)$ reads
\begin{eqnarray}
&&{}V_{\lambda_q\lambda_{\bar{q}}}^{c_1c_2,\,\mu\nu}(q_1,q_2,k_1,k_2)
=-\frac{g_s^2}{2}\,
\delta^{c_1c_2}\,\bar{u}_{\lambda_q}(k_1) \nonumber \\
&&\biggl(\gamma^{\nu}\frac{\hat{q}_{1}-\hat{k}_{1}-m}
{(q_1-k_1)^2-m^2}\gamma^{\mu} 
-\gamma^{\mu}\frac{\hat{q}_{1}-
\hat{k}_{2}+m}{(q_1-k_2)^2-m^2}\gamma^{\nu}\biggr)v_{\lambda_{\bar{q}}}(k_2)\nonumber.\\
 \label{vector_tensor}
\end{eqnarray}

In the present calculations we use the GJR08 set of collinear gluon
distributions \cite{GJR}. In the analogy to the CEP of Higgs boson,
where renormalization and factorization scales  
$\mu^{2} = \mu_{R}^2= \mu_{F}^2 = M_{H}^{2}$ are preferred, we 
take $\mu^{2} = M_{c \bar c}^2$. Absorption 
corrections to the bare $p p \rightarrow p (q \bar q) p$ amplitude,
which are necessary to be taken into account (to ensure exclusivity of 
the process), are included approximately by multiplying the cross section 
by the gap survival factors $S_G = 0.1$ for RHIC and $S_G = 0.03$ for 
the LHC energy.
More details about exclusive production of heavy quarks can be found in
our original paper \cite{MPS2010}.
Let us come now to presentation of our results.

In Fig.~\ref{fig:dsig_dy_exc} we show rapidity distribution
of $c$ quarks from the exclusive mechanism (solid lines) shown already in 
Fig.~\ref{fig:CED_mechanism} . We show the results
for LO (upper curves) and NLO (lower curves)
collinear gluon distributions \cite{GJR}.
We observe large difference of results for LO and NLO gluon distributions
especially at LHC.
For comparison we show the contribution of inclusive central diffractive
component discussed in detail in \cite{LMS11}.
In this calculation we have included gap survival factors
$S_G$ = 0.1 for $\sqrt{s}$ = 500 GeV and $S_G$ = 0.03 for $\sqrt{s}$ =
14 TeV.
The cross section for the exclusive mechanism is similar to that 
for the inclusive central diffractive mechanism. 
The exclusive production starts to dominate only at large $c$ quark 
rapidities. Therefore a measurement of the cross
section with double (both side) rapidity gaps may be not sufficient
to single out the exclusive mechanism. Clearly other cuts would be
necessary.

\begin{figure} [!thb]
\begin{center}
\includegraphics[width=6cm]{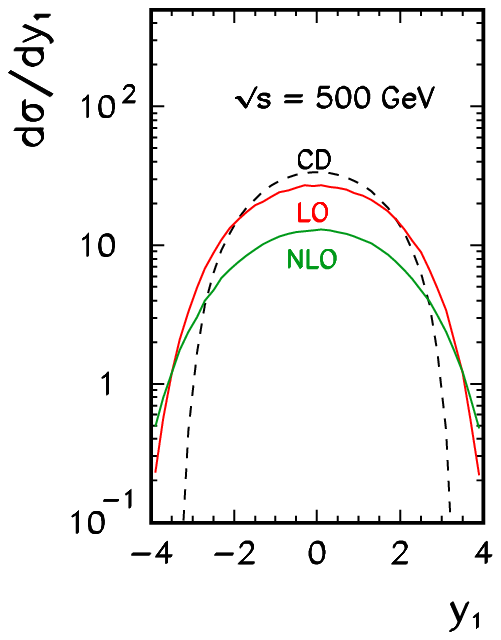}
\includegraphics[width=6cm]{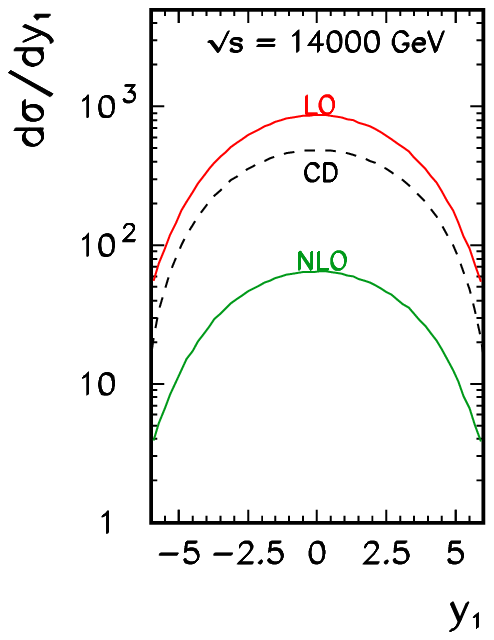}
\end{center}
\caption{\small Distributions in rapidity 
of $c$ quark/antiquark for the exclusive component 
at $\sqrt{s}$ = 500 GeV (left panel)
and $\sqrt{s}$ = 14 TeV (right panel).
GJR08 collinear gluon distributions were used to obtain
the unintegrated gluon distribution according to the KMR prescription.
For comparison we show the inclusive central diffractive contribution (dashed line).
\label{fig:dsig_dy_exc}
}
\end{figure}

Distributions in the $c$ quark ($ \bar c$ antiquark) 
transverse momentum are shown in Fig.~\ref{fig:dsig_dpt_exc}. At RHIC 
energy distributions for both mechanisms have very similar shape. 
However, at LHC nominal energy we observe that inclusive central 
diffractive component extends to higher transverse momentum than that 
for the exclusive central diffractive one. 
In order to identify the exclusive component a much more precise
analysis of kinematical correlations between quark and antiquark is needed.
A detailed Monte Carlo studies of final states of both mechanisms could 
help to find a criterion to separate experimentally the two dynamically
different components.

\begin{figure} [!thb]
\begin{center}
\includegraphics[width=6cm]{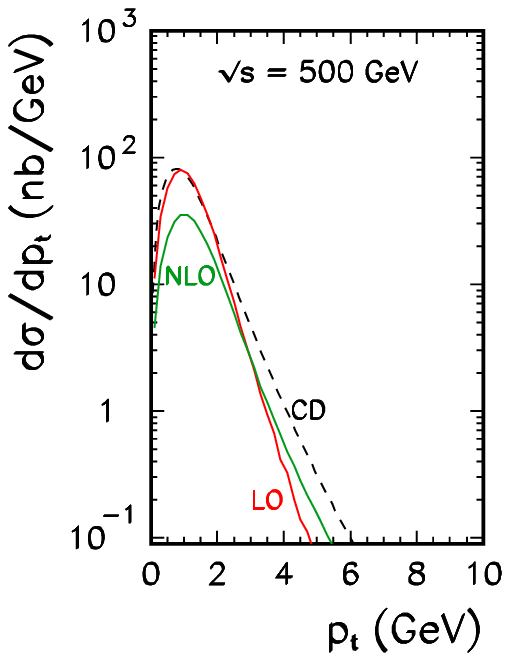}
\includegraphics[width=6cm]{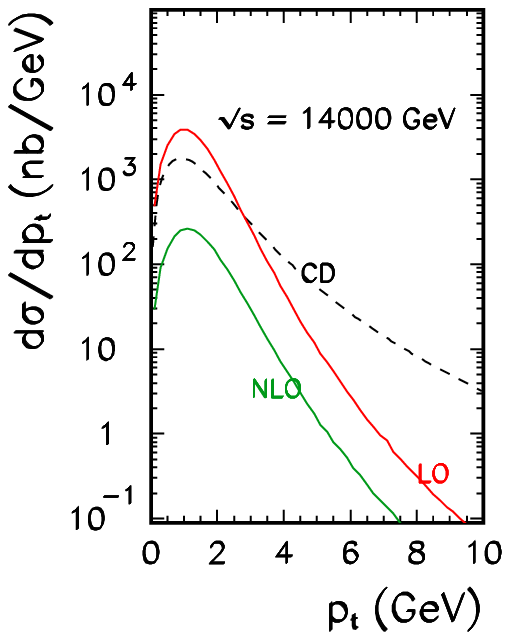}
\end{center}
\caption{\small Distributions in transverse momentum 
of $c$ quark/antiquark for the exclusive component 
at $\sqrt{s}$ = 500 GeV (left panel)
and $\sqrt{s}$ = 14 TeV (right panel).
GJR08 collinear gluon distributions were used to obtain
the unintegrated gluon distribution.
For comparison we show the inclusive central diffractive contribution 
(dashed line).
\label{fig:dsig_dpt_exc}
}
\end{figure}

\section{Double parton scattering production of $c \bar c c \bar c$}

\subsection{Framework}

The double-parton scattering has been recognized long ago. 
Several estimates of the cross section for different processes have been
presented in recent years.
In the present analysis we discuss production of $(c \bar c) (c \bar c)$
four-parton final state which has not been discussed so
far but is particularly interesting especially in 
the context of experiments being carried out at LHC.


The double-parton scattering formalism proposed so far assumes two
single-parton scatterings. In a simple probabilistic picture the cross
section for double-parton scattering can be written as:
\begin{equation}
\sigma^{DPS}(p p \to c \bar c c \bar c X) = \frac{1}{2 \sigma_{eff}}
\sigma^{SPS}(p p \to c \bar c X_1) \cdot \sigma^{SPS}(p p \to c \bar c X_2).
\label{basic_formula}
\end{equation}
This formula assumes that the two subprocesses are not correlated and do
not interfere.
At low energies one has to include parton momentum conservation
i.e. extra limitations: $x_1+x_3 <$ 1 and $x_2+x_4 <$ 1, where $x_1$ and $x_3$
are longitudinal momentum fractions of gluons emitted from one proton
and $x_2$ and $x_4$
their counterpairs for gluons emitted from the second proton. 
Experimental data provide an estimate of $\sigma_{eff}$
in the denominator of formula (\ref{basic_formula}). In our analysis we
take $\sigma_{eff}$ = 15 mb.

\begin{figure}[!h]
\begin{center}
\includegraphics[width=5cm]{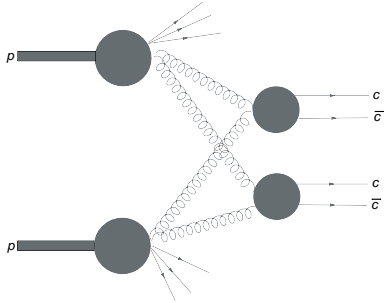}
\end{center}
   \caption{
\small Mechanism of $(c \bar c) (c \bar c)$ production via double-parton 
scattering. 
}
 \label{fig:diagram}
\end{figure}

The simple formula (\ref{basic_formula}) can be generalized to include 
differential distributions. In the same approximation 
differential distribution can be written as
\begin{equation}
\frac{d \sigma}{d y_1 d y_2 d^2 p_{1t} d y_3 d y_4 d^2 p_{2t}}  \\ =
\frac{1}{ 2 \sigma_{eff} }
\frac{ d \sigma } {d y_1 d y_2 d^2 p_{1t}} \cdot
\frac{ d \sigma } {d y_3 d y_4 d^2 p_{2t}} 
\label{differential_distribution}
\end{equation}
which reproduces formula (\ref{basic_formula}).
This cross section is formally differential in 8 dimensions but can be 
easily reduced to 7 dimensions noting that physics of unpolarized
scattering cannot depend on azimuthal angle of the pair or on azimuthal angle of one of 
the produced $c$ ($\bar c$) quark (antiquark).
The differential distributions for each single scattering step can be written in terms
of collinear gluon distributions with longitudinal momentum fractions
$x_1$, $x_2$, $x_3$ and $x_4$ expressed in terms of rapidities $y_1$, $y_2$, $y_3$,
$y_4$ and transverse momenta of quark (or antiquark) for each step. 

A slightly more general formula for the cross section can be written formally 
in terms of double-parton distributions (dPDF), e.g. $F_{gg}$, $F_{qq}$, etc. 
In the case of heavy quark (antiquark) production at high energies:
\begin{eqnarray}
d \sigma^{DPS} &=& \frac{1}{2 \sigma_{eff}}
F_{gg}(x_1,x_3,\mu_1^2,\mu_2^2) F_{gg}(x_{2},x_{4},\mu_1^2,\mu_2^2)
\nonumber \\
&&d \sigma_{gg \to c \bar c}(x_1,x_2,\mu_1^2)
d \sigma_{gg \to c \bar c}(x_3,x_4,\mu_2^2) \; dx_1 dx_2 dx_3 dx_4 \, .
\label{cs_via_doublePDFs}
\end{eqnarray}
%

The double-parton distributions in Eq.(\ref{cs_via_doublePDFs})
are not well known. Usually one assumes the factorized form and
expresses them via standard distributions for SPS.
Even if factorization is valid at some scale, QCD evolution leads
to a factorization breaking \cite{LMS11_DPS}.

In this presentation we shall apply the commonly
used factorized model. Some refinements are presented in 
\cite{LMS11_DPS}. 

\subsection{Results}

In Fig.~\ref{fig:single_vs_double_LO} we
compare cross sections for single $c \bar c$ and DPS $c \bar c c \bar c$
production as a function of $p p$ center-of-mass energy. At low energies the 
cross section for $c \bar c$ is much larger. For reference we show the 
proton-proton total cross section as a function of energy.
At low energy the $c \bar c$ or $ c \bar c c \bar c$ cross sections are much
smaller than the total cross section. At higher energies the contributions
approach the parametrized total cross section.
This shows that inclusion of
unitarity effect and/or saturation of parton distributions may be necessary.
At LHC energies the cross section for both terms becomes comparable.
This is a new situation when the double-parton scattering gives a large 
contribution to inclusive charm production. This issue was not discussed
so far in the literature.

\begin{figure}[!h]
\begin{center}
\includegraphics[width=5cm]{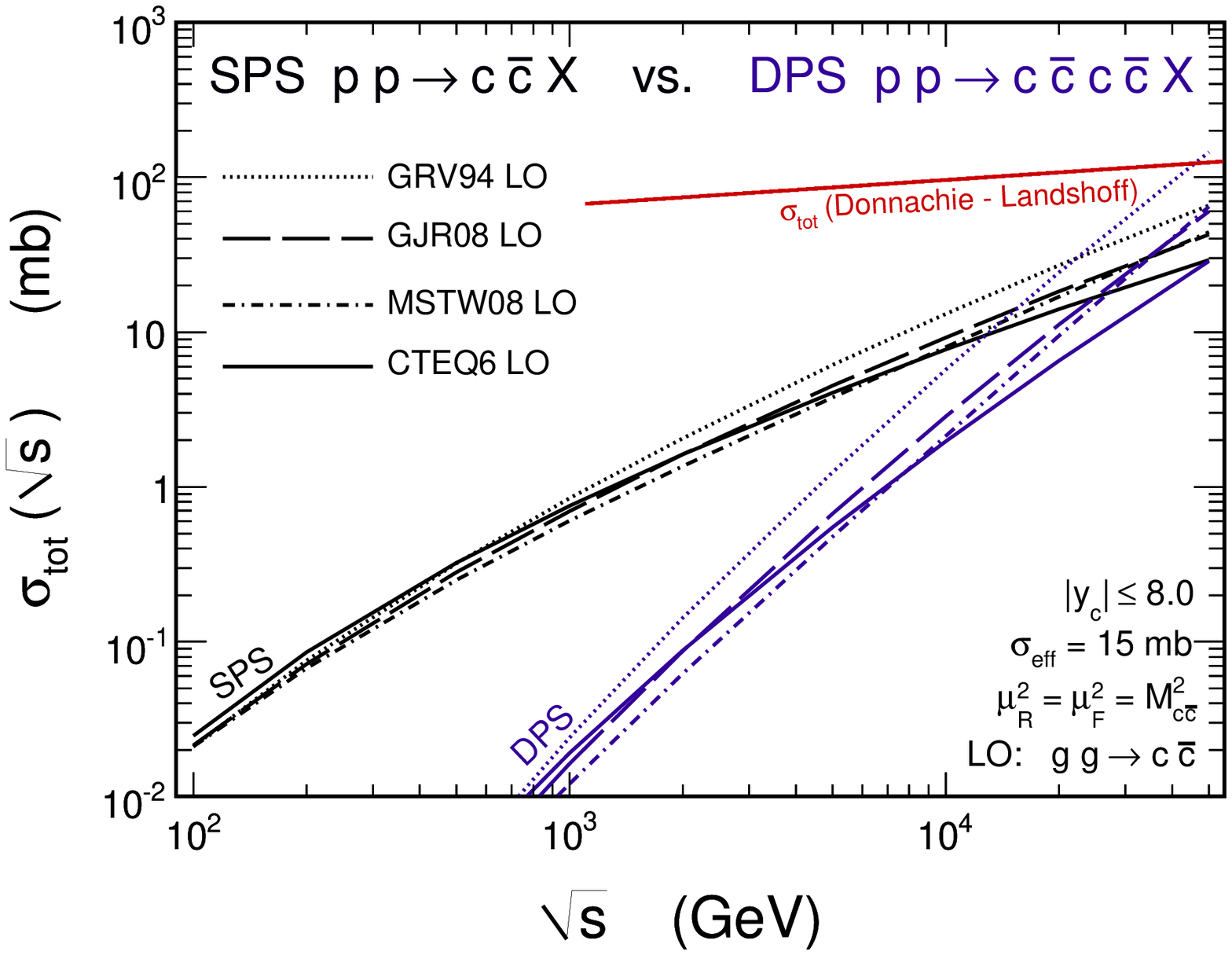}
\includegraphics[width=5cm]{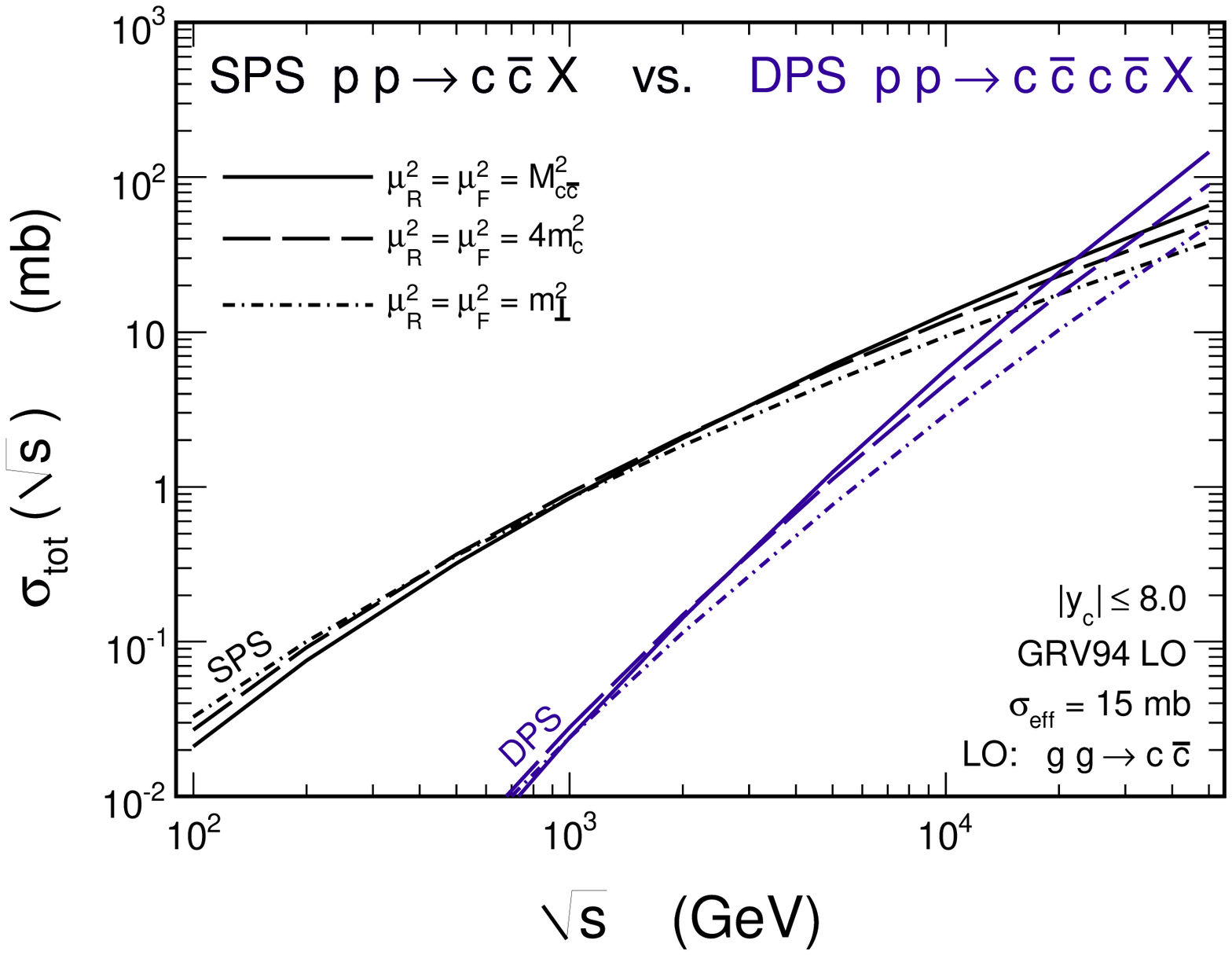}
\end{center}
   \caption{
\small Total LO cross section for single-parton and double-parton
scattering as a function of center-of-mass energy (left panel) and 
uncertainties due to the choice of (factorization, renormalization) scales (right panel). 
We show in addition a parametrization of the total cross section in the left panel.
}
 \label{fig:single_vs_double_LO}
\end{figure}

As an example in Fig.~\ref{fig:double_single1} we present
single $c$ ($\bar c$) distributions. Within approximations discussed
here the single-parton distributions are identical in shape 
for $c \bar c$ and $c \bar c c \bar c$.
This means that double-scattering contribution produces naturally an
extra energy-dependent $K$-factor
to be contrasted with approximately energy-independent $K$-factor 
due to higher-order corrections. A strong dependence on 
the factorization and renormalization scales can be observed.

\begin{figure}[!h]
\begin{center}
\includegraphics[width=5cm]{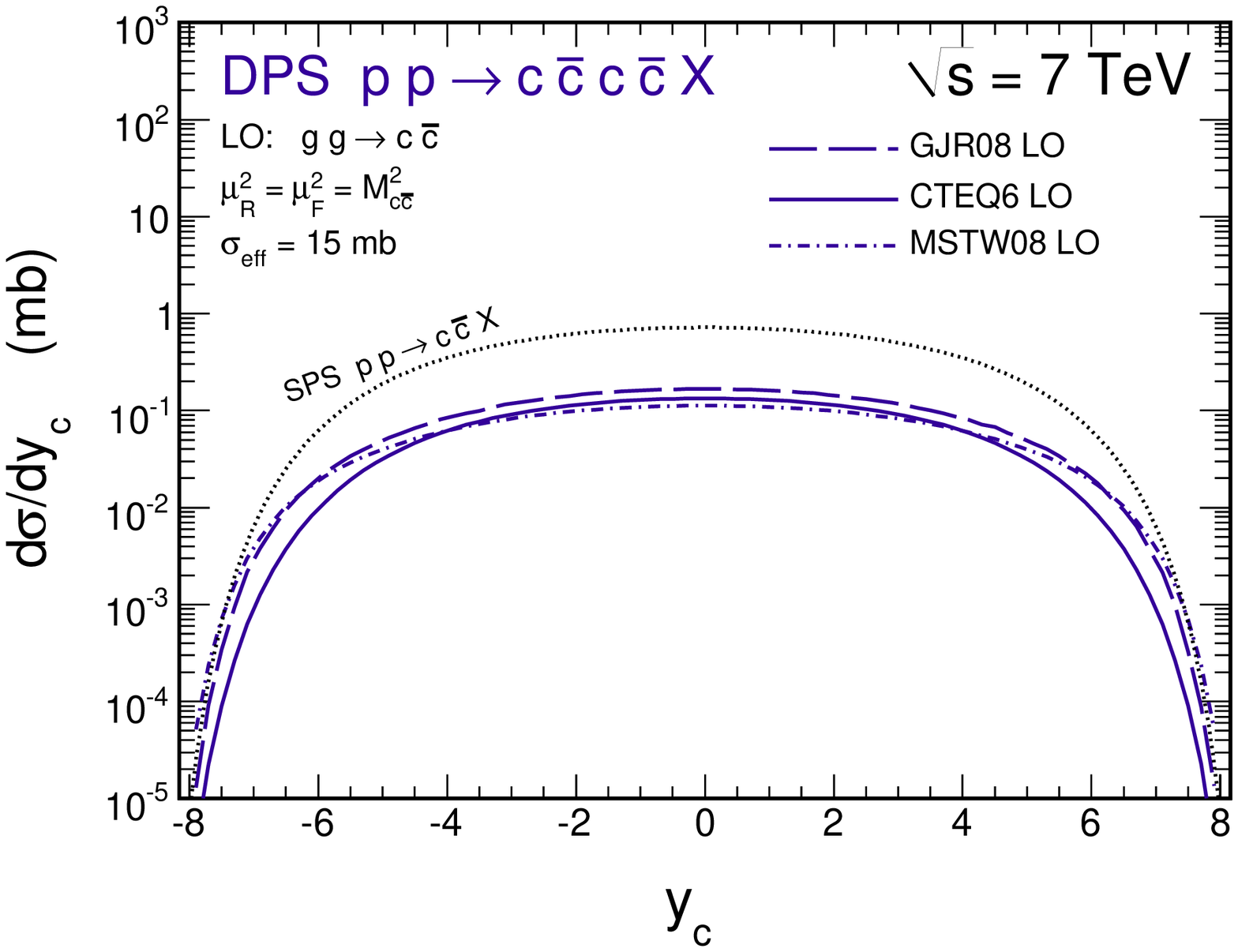}
\includegraphics[width=5cm]{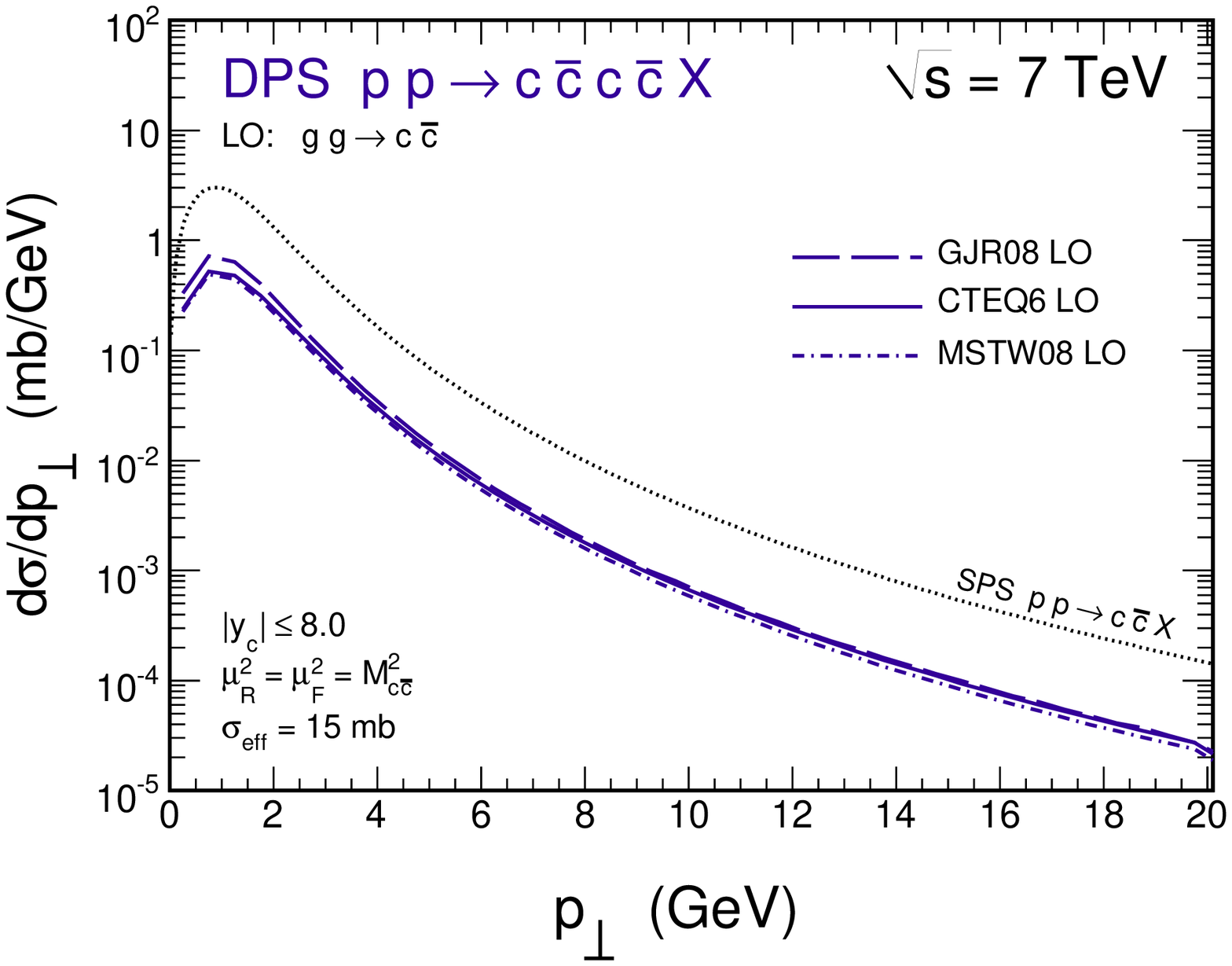}
\includegraphics[width=5cm]{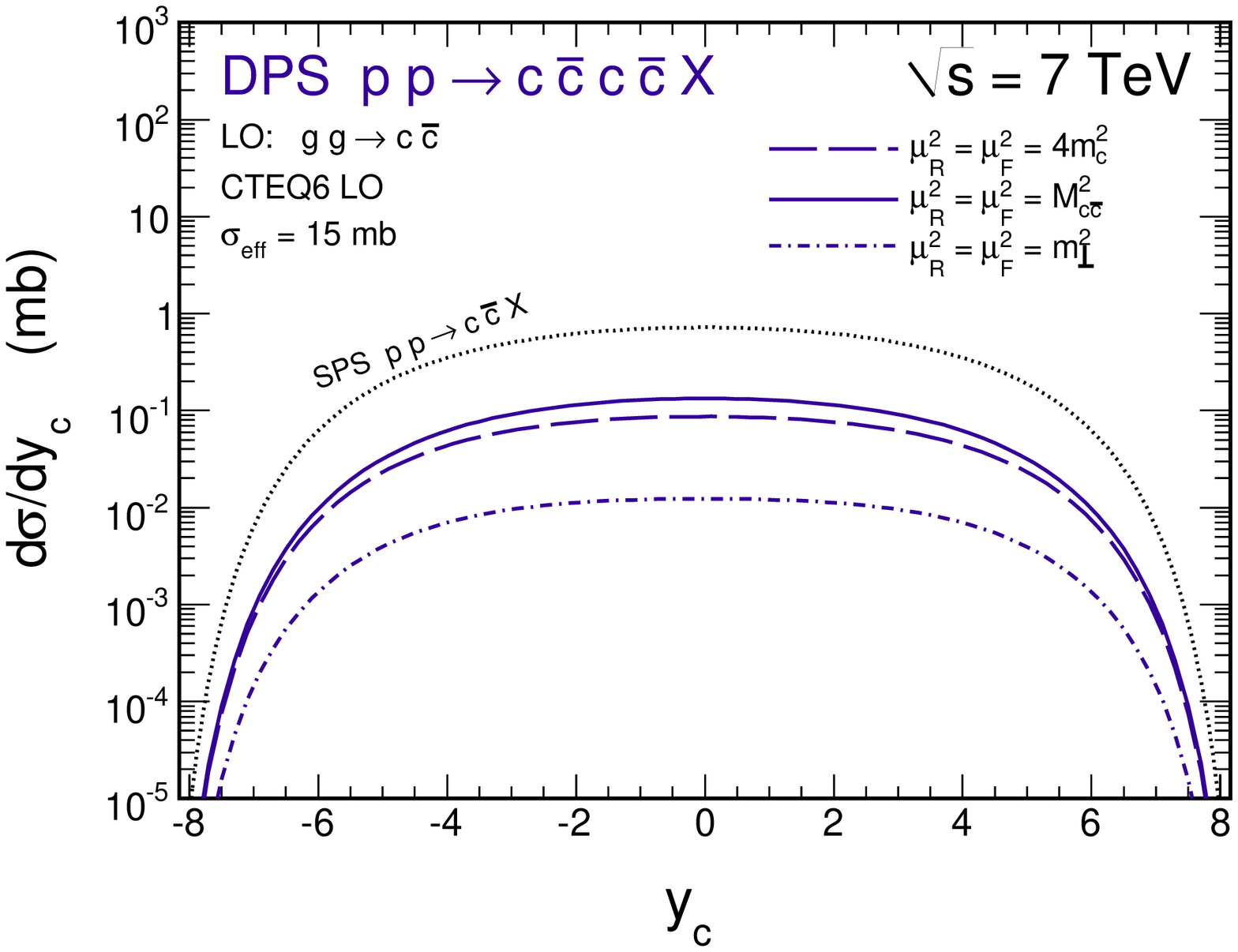}
\includegraphics[width=5cm]{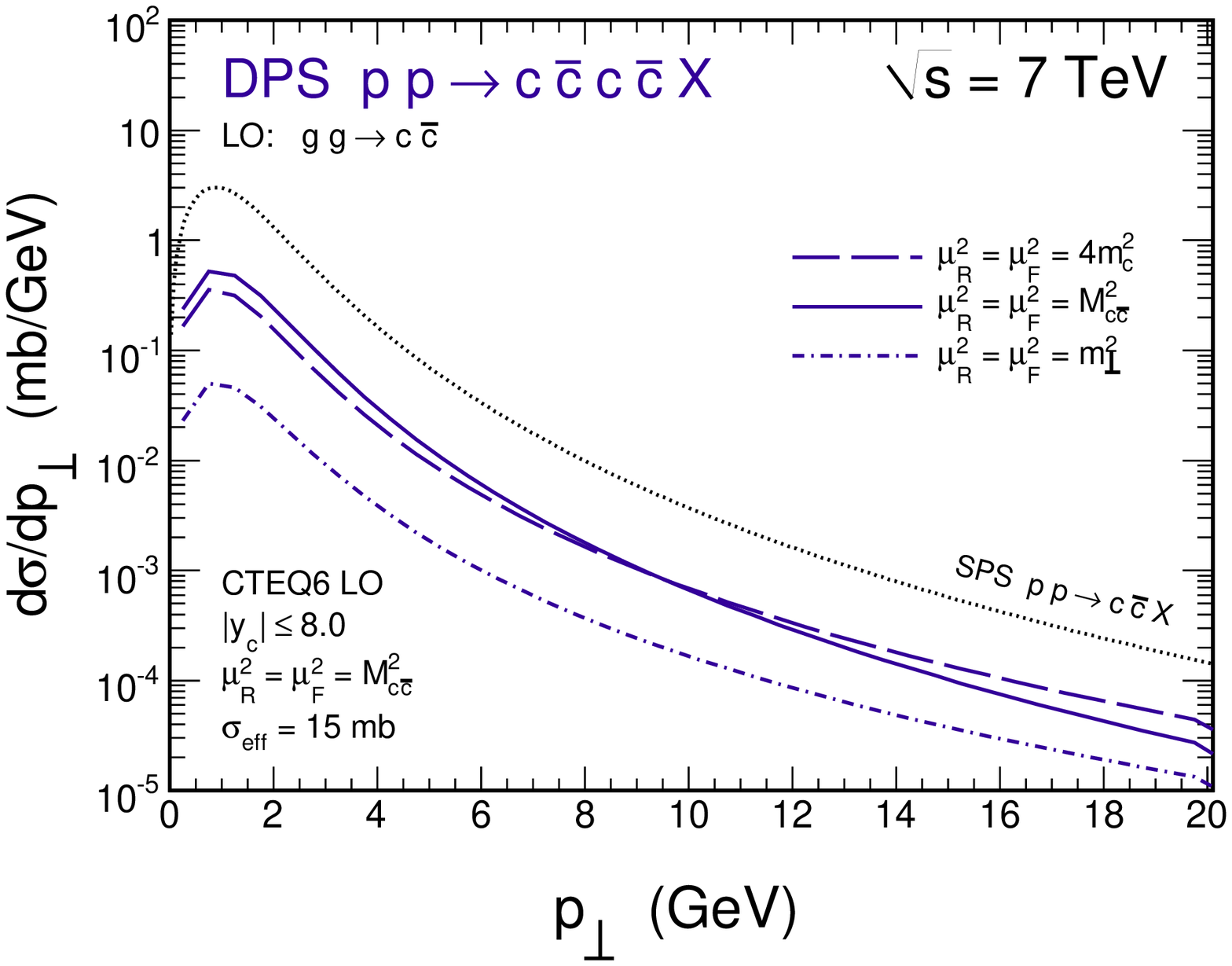}
\end{center}
   \caption{
\small Distribution in rapidity (left upper panel) and transverse
momentum (right upper panel) for different UGDFs and associated
ucertainties related to the choice of renormalization and factorization 
scales (lower panels) for $c$ or $\bar{c}$ quarks  at $\sqrt{s}$ = 7 TeV. 
}
 \label{fig:double_single1}
\end{figure}


So far we have discussed only single particle spectra of $c$ or $\bar c$.
A better test of DPS could be done by studying correlation observables.
The correlations between $c$ and $\bar c$ has been studied e.g. in \cite{LS06}.
In Fig.~\ref{fig:double_correlations_1} we show distribution in the
difference of $c$ and $\bar c$ rapidities
(left panel) as well as in the $c \bar c$ invariant mass $M_{c\bar c}$ 
(right panel). We show both cases: when $c \bar c$ are emitted
in the same parton scattering ($c_1\bar c_2$ or $c_3\bar c_4$) and 
when they are emitted from different parton scatterings 
($c_1\bar c_4$ or $c_2\bar c_3$). 
We observe a long tail for large rapidity difference as well as at large
invariant masses of $c \bar c$.
Such distributions for quarks and antiquarks cannot be directly measured.
Instead their counterparts for mesons can be studied. This was discussed
in more detail in our recent original paper \cite{LMS11_DPS}.



\begin{figure}[!h]
\begin{center}
\includegraphics[width=5.0cm]{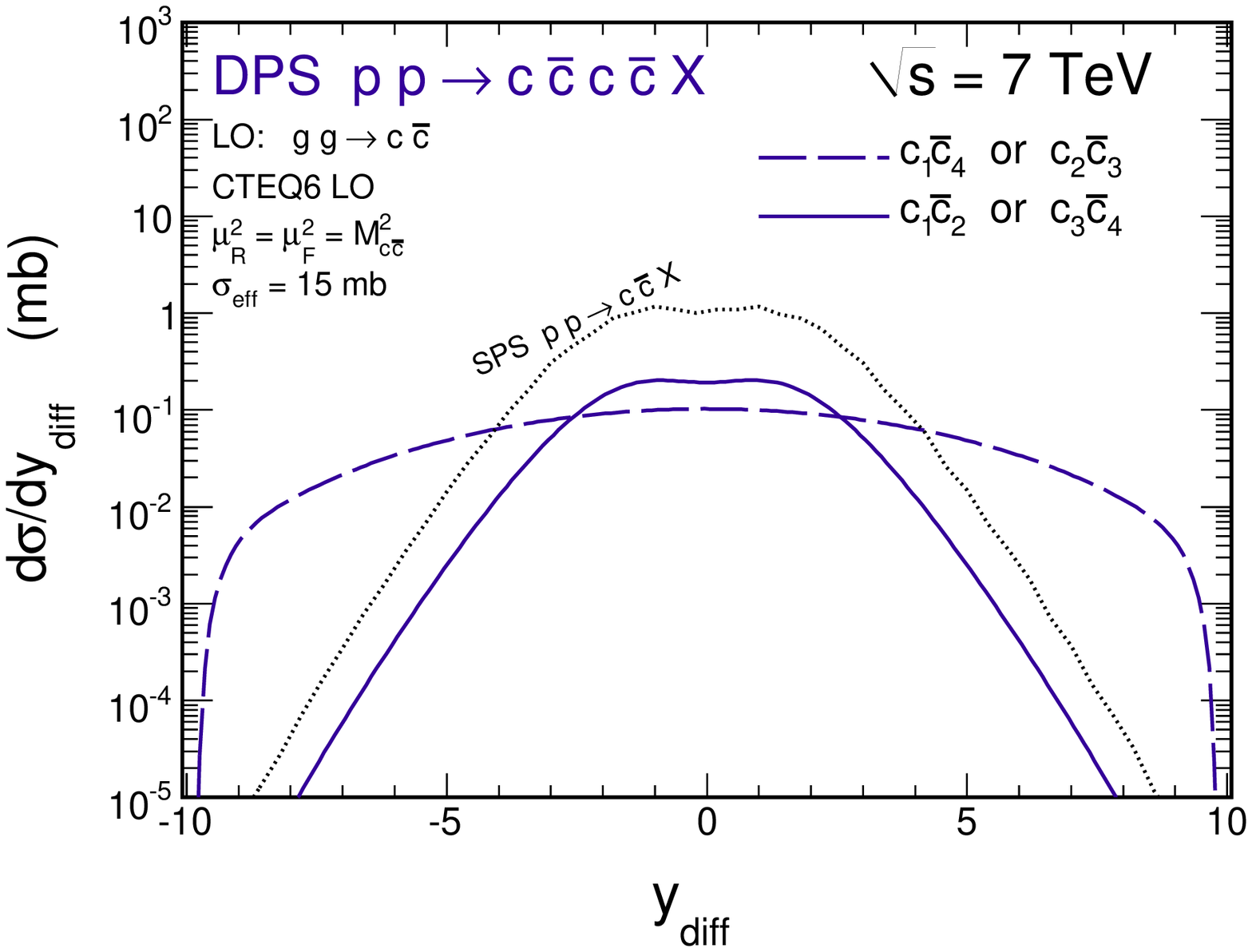}
\includegraphics[width=5.0cm]{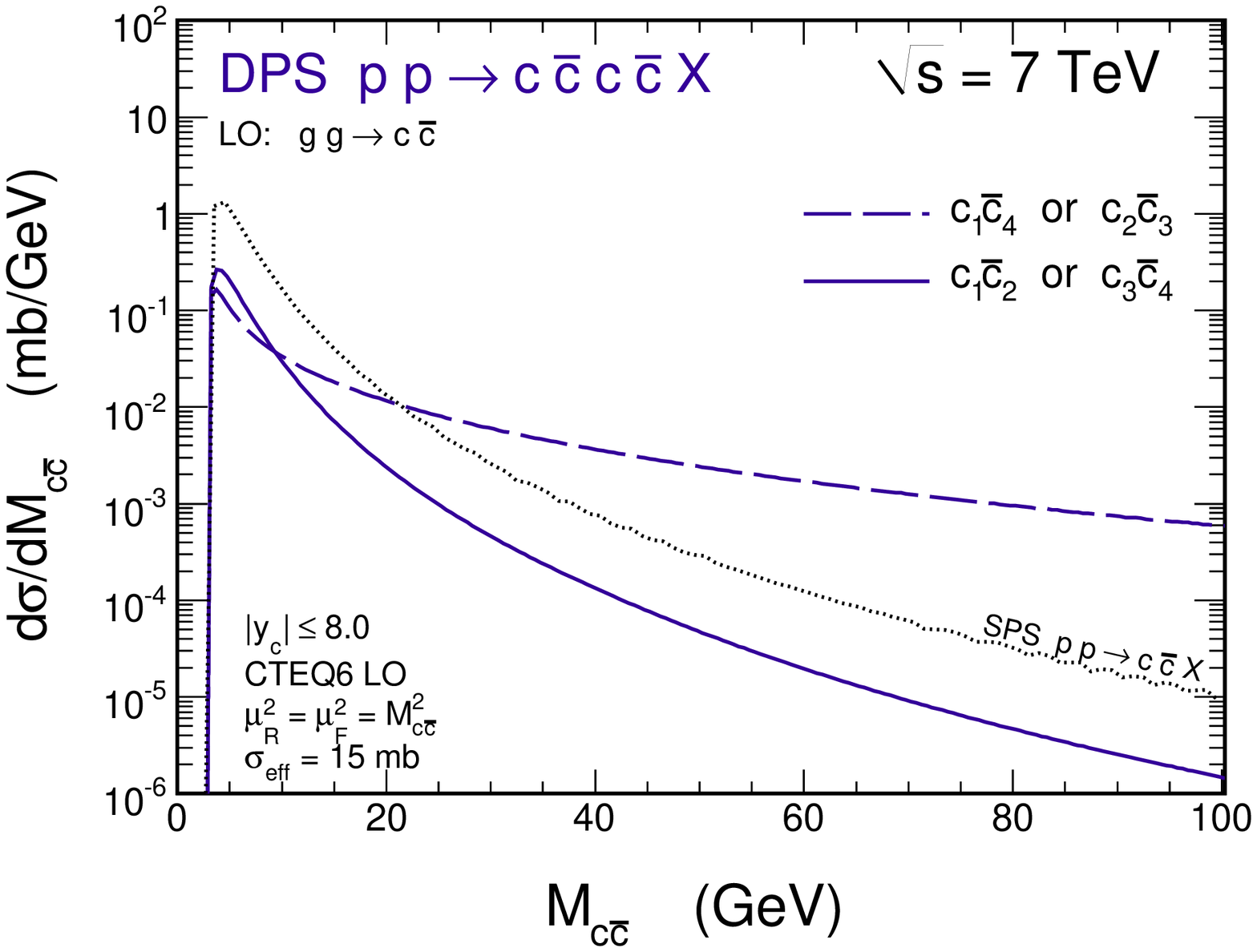}
\end{center}
   \caption{
\small Distribution in rapidity difference (left panel) and in invariant
mass of the $c\bar{c}$ pair (right panel) at $\sqrt{s}$ = 7 TeV.
}
\label{fig:double_correlations_1}
\end{figure}

As the last example in Fig.~\ref{fig:double_correlations_2} we present 
distribution in the transverse momentum of the $c \bar c$ pair
$|\overrightarrow{p_{\perp c\bar c}}|$, 
where $\overrightarrow{p_{\perp c\bar c}} = \overrightarrow{p_{\perp c}}
+ \overrightarrow{p_{\perp \bar c}}$. For comparison this is a Dirac
delta function in the leading-order approximation to $c \bar c$
production. In contrast, double-parton scattering mechanism gives a
broad distribution extending to large transverse
momenta. NLO corrections obviously destroy the $\delta$-like
leading-order correlation. 
Similar distributions for $D \bar D$ 
seem useful observables to identify the DPS contributions \cite{LMS11_DPS}.

So far we have calculated cross section in a simple
leading-order approach. A better approximation would be to include
multiple gluon emissions. This can be done e.g. in soft gluon resummation
or in the $k_t$-factorization approach. For example
the second approach does not lead to large changes in shape of neither 
distributions in rapidity nor of distributions in transverse momentum 
of $c$ ($\bar c$) (see e.g. \cite{LS06}) compared to the collinear
approach. It is expected, however, to change distributions in transverse
momentum of $c \bar c$ or in azimuthal angle between $c$ and $ \bar c$ 
\cite{LS06}. 


 
\begin{figure}[!h]
\begin{center}
\includegraphics[width=5cm]{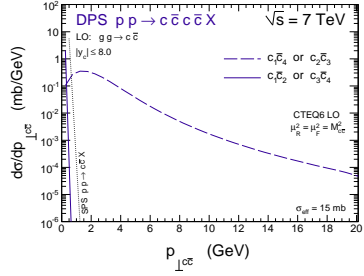}
\end{center}
   \caption{
\small Distribution in transverse momentum of $c\bar{c}$ pairs from the
same parton scattering and from different parton scatterings at 
$\sqrt{s}$ = 7 TeV. 
}
 \label{fig:double_correlations_2}
\end{figure}



\section{Conclusions}

We have calculated inclusive spectra of nonphotonic
electrons/positrons for RHIC energy in the framework
of the $k_t$-factorization. We have concentrated on 
the dominant gluon-gluon fusion mechanism and used
unintegrated gluon distribution functions
from the literature.
Special emphasis was devoted
to the Kwieci\'nski unintegrated gluon (parton)
distributions. In this formalism, using unintegrated
quark and antiquark distributions, one can calculate
also the quark-antiquark annihilation process
including transverse momenta of initial
quarks/antiquarks. 

When calculating spectra of charmed ($D$, $D^*$) and 
bottom ($B$, $B^*$) mesons we have used Peterson
and Braaten et al. fragmentation functions.
We have used recent fits to the CLEO and BABAR 
collaborations data for decay functions of heavy mesons.

Our results have been compared with experimental data measured
recently by the PHENIX and STAR collaborations at RHIC. A reasonable 
description of the data at large transverse momenta of 
electrons/positrons has been achieved. We have discussed
uncertainties related to the choice of the factorization
and renormalization scales as well as
those related to the fragmentation process. Although the uncertainty 
bands are rather large, 
there seems to be a missing strength at lower electron/positron 
transverse momenta.

We have discussed also correlations of charmed mesons
and dielectrons at the energy of recent RHIC experiments. We have
calculated the spectra in dielectron invariant mass, in azimuthal angle 
between electron and positron as well as the distribution in
transverse momentum of the pair.
The uncertainties due to the choice of UGDFs, choice of the
factorization and renormalization scales, have been discussed.
We have obtained good description of the dielectron invariant mass
distribution measured recently by the PHENIX collaboration at RHIC. 

At RHIC the contribution of electrons from Drell-Yan processes is only 
slightly smaller than that from the semileptonic decays. The
distributions in azimuthal angle between electron and positron and in 
the transverse momentum of the dielectron pair from both processes are 
rather similar. 
It was found that the distribution in azimuthal angle strongly depends
on dielectron invariant mass.

We have also included exclusive central-diffractive contribution
discussed recently in the literature.
At the rather low RHIC energy it gives, however, a very small contribution
to the cross section and can be safely ignored. 

The QED double-elastic, double-inelastic, elastic-inelastic and 
inelastic-elastic processes give individually
rather small contribution but when added together are not negligible
especially at low dielectron invariant masses where some strength is 
missing.


The exclusive production of $c \bar c$ pairs is interesting by itself.
We have discussed corresponding formalism as well as some results
for RHIC and LHC energies. However, experimental identification of 
the mechanism may be not easy as the final hadronic state is more
complicated and will compete with inclusive central diffractive
production of $c \bar c$.

We have discussed also production of two pairs of $c \bar c$.
We have found very quickly rising cross section for the two-pair
production as a function of center-of-mass energy. The two-pair
production must therefore give a sizeable contribution to inclusive
charm production. This point requires further studies.

We have discussed some promissing observables which seem useful
in identifying the DPS production of two pairs of $c \bar c$.
In Ref.\cite{LMS11_DPS} we have considered also corresponding 
observables for charmed mesons. Another option would be to study
production of the same-sign charged leptons.
We expect that semileptonic decays are the main source for semi-hard
muons or electrons. Furthermore this contribution can, in principle, be
separated experimentally by taking into account that the secondary
vertices are shifted with respect to the primary ones. This should allow
a separation of the semileptonic "signal" from other possible sources of
dilepton continuum.

\vspace{0.5cm}


\end{document}